\documentclass[journal]{vgtc}         

\vgtcinsertpkg
 
\ifpdf
  \pdfoutput=1\relax                   
  \usepackage{graphicx}                
  \DeclareGraphicsExtensions{.pdf,.png,.jpg,.jpeg} 
\else
  \usepackage{graphicx}                
  \DeclareGraphicsExtensions{.eps}     
\fi%

\graphicspath{{figures/}{figs/}{images/}{./}} 

\usepackage{microtype}                 
\PassOptionsToPackage{warn}{textcomp}  
\usepackage{textcomp}                  
\usepackage{mathptmx}                  
\usepackage{times}                     
\usepackage{cite}                      
\usepackage{tabu}                      
\usepackage{booktabs}                  
\usepackage{tcolorbox}
\usepackage{lipsum}
\usepackage{tikz}
\usepackage{balance}
\usepackage{xcolor}
\usepackage{color}
\usepackage{xspace}
\usepackage{mathptmx}
\usepackage{graphicx,stringenc}
\usepackage{xparse,l3regex}
\usepackage{times}
\usepackage{amsmath}
\usepackage{amssymb}

\usepackage{amsthm}
\usepackage{amsfonts}
\usepackage{subcaption}
\usepackage{boxedminipage}
\usepackage{algorithmic}
\usepackage[algoruled,linesnumbered]{algorithm2e}
\usepackage{comment}
\usepackage{url}
\usepackage{enumitem}
\usepackage{xfrac}
\usepackage{tabularx}
\usepackage{multirow}
\usepackage{array}
\usepackage{bigstrut}
\usepackage[usestackEOL]{stackengine}
\usepackage{bibunits}
\usepackage{wrapfig}
\usepackage{booktabs} 
\usepackage{quotes}
\usepackage{float}
\usepackage{longtable,booktabs}
\usepackage{makecell}
\usepackage{tikz}
\usepackage{calc}
\usepackage{tabularx,booktabs}
\usepackage{fancybox}

\ExplSyntaxOn
\tl_const:Nn \c_coloremoji_dir_tl { ./emoji_images/lowres }
\tl_new:N \l_coloremoji_input_tl

\cs_new_protected:Npn \coloremoji_unicode:nn #1 #2
 {
  \includegraphics[#1]{\c_coloremoji_dir_tl / \int_to_Hex:n{`#2}.pdf}
 }

\cs_new_protected:Npn \coloremoji_eightbit:nn #1 #2
 {
  \StringEncodingConvert \l_coloremoji_input_tl
    { \tl_to_str:n { #2 } } { utf8 } { utf32 }
  \tl_set:Nx \l_coloremoji_input_tl { \pdfescapehex\expandafter{\l_coloremoji_input_tl} }
  \regex_replace_once:nnN { \A 0* } { } \l_coloremoji_input_tl
  \includegraphics[#1]{\c_coloremoji_dir_tl / \l_coloremoji_input_tl.pdf}
 }
 
\NewDocumentCommand{\coloremoji}{O{}m}
 {
  \bool_if:nTF { \xetex_if_engine_p: || \luatex_if_engine_p: }
   {
    \coloremoji_unicode:nn { #1 } { #2 }
   }
  {
   \coloremoji_eightbit:nn { #1 } { #2 }
  }
 }

\ExplSyntaxOff

\cornersize{0.7}

\newcommand{\xvar}[1]{\textsf{#1}}

\newcommand{\xvbox}[2]{\makebox[#1][l]{#2}}

\SetAlFnt{\footnotesize}

\SetAlCapSty{xAlCapSty}

\SetCommentSty{xCommentSty}

\SetNlSty{mynlfont}{}{} 

\LinesNumbered

\SetSideCommentRight

\DontPrintSemicolon

\RestyleAlgo{algoruled}
 
\onlineid{1237}

\vgtccategory{Research}
\vgtcpapertype{technique}

\title{\algoname: Generating and Visualizing Machine Learning Model \\   
Summary to Understand Data and Classifiers Together}

\author{Gromit Yeuk-Yin Chan, Luis Gustavo Nonato, Brian Barr, Enrico Bertini, Cl\'{a}udio T. Silva}
\authorfooter{
\item
Gromit Yeuk-Yin Chan, Enrico Bertini and Cl\'{a}udio T. Silva are with New York University. 
E-mail: \href{mailto:gromit.chan@nyu.edu}{gromit.chan},\href{mailto:enrico.bertini@nyu.edu}{enrico.bertini},\href{mailto:csilva@nyu.edu}{csilva}@nyu.edu.
\item
Luis Gustavo Nonato is with University of S{\~a}o Paulo.
E-mail: \href{mailto:gnonato@icmc.usp.br}{gnonato@icmc.usp.br}.
\item
Brian Barr is with Capital One.
E-mail: \href{mailto:brian.barr@capitalone.com}{brian.barr@capitalone.com}.
}

\shortauthortitle{Chan \MakeLowercase{\textit{et al.}}: \algoname: Generating and Visualizing Machine Learning Model Summary to Understand Data and Classifiers Together}

\abstract{
With the increasing sophistication of machine learning models, there are growing trends of developing model explanation techniques that focus on only one instance (local explanation) to ensure faithfulness to the original model.
While these techniques provide accurate model interpretability on various data primitive (e.g., tabular, image, or text), a holistic Explainable Artificial Intelligence (XAI) experience also requires a global explanation of the model and dataset to enable sensemaking in different granularity. 
Thus, there is a vast potential in synergizing the model explanation and visual analytics approaches. 
In this paper, we present \algoname, an interactive algorithm to construct an optimal global overview of the model and data behavior by summarizing the local explanations using information theory. The result (i.e., an explanation summary) does not require additional learning models, restrictions of data primitives, or the knowledge of machine learning from the users. 
We also design \systemname, an interactive visual analytics system to demonstrate how the explanation summary connects the dots in various XAI tasks from a global overview to local inspections. 
We present three usage scenarios regarding tabular, image, and text classifications to illustrate how to generalize model interpretability of different data.
Our experiments show that our approaches: (1) provides a better explanation summary compared to a straightforward information-theoretic summarization and (2) achieves a significant speedup in the end-to-end data modeling pipeline.
} 


\CCScatlist{ 
 \CCScat{K.6.1}{Management of Computing and Information Systems}%
{Project and People Management}{Life Cycle};
 \CCScat{K.7.m}{The Computing Profession}{Miscellaneous}{Ethics}
}

\teaser{
  \centering
  \includegraphics[width=\linewidth]{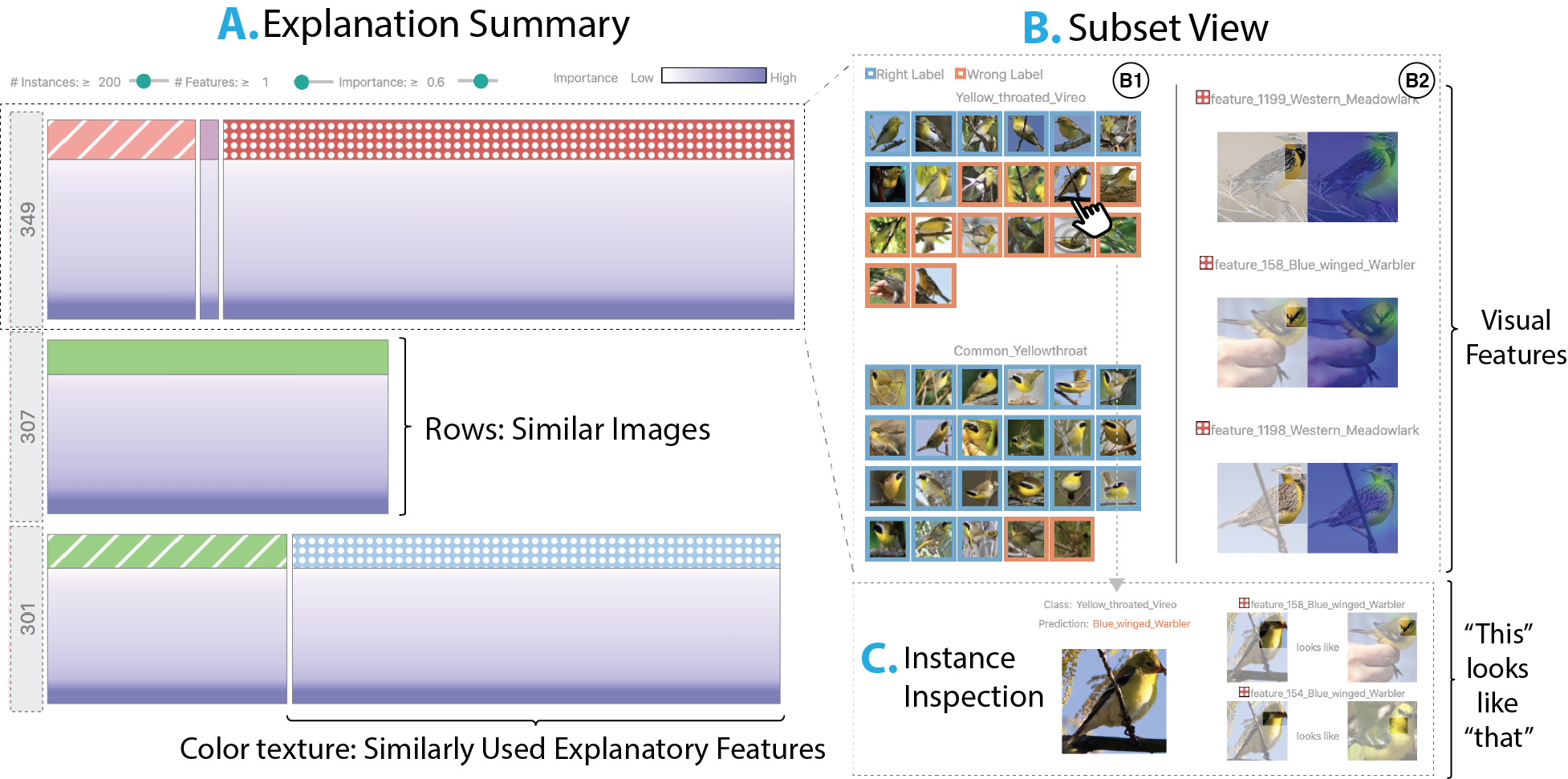}
  \caption{
    \textbf{A.} A model explanation summary generated from a deep neural network that classifies birds images into 200 species.
      It summarizes groups of similar images (rows) and groups of common visual representations (color texture) used by different groups to explain the predictions. 
      \textbf{B.} A subset view is shown to display the details inside a group. For example, here reveals the model learns the ``yellowish'' patches around the birds' upper bodies (right) to classify the species related to ``yellowthroat'' (left).
      \textbf{C.} Users can understand the decision logic of individual examples in full details. Here is an example of the wrong prediction on a yellow throated vimeo since its peck and neck are similar to the features learned from a blue winged warbler in the model.
  }
	\label{fig:teaser}
}

\renewcommand{\paragraph}[1]{\vspace{0pt} #1}

\newcommand{\hidecomment}[1]{}
\newcommand{\savespace}[1]{}

\newcommand{\systemname}{\textsc{Melody UI}\xspace}
\newcommand{\algoname}{\textsc{Melody}\xspace}

\newcommand{\etal}{\textit{et al.}\xspace}

\newcommand{\clabel}[1]{\textcircled{\tiny{\textbf{#1}}}}

\definecolor{darkpastelgreen}{rgb}{0.01, 0.75, 0.24}
\definecolor{lavenderblush}{rgb}{1.0, 0.94, 0.96}
\definecolor{amber(sae/ece)}{rgb}{1.0, 0.49, 0.0}
\definecolor{lavender(web)}{rgb}{0.9, 0.9, 0.98}

\newtcbox{\boxcolor}[1][lavenderblush]{on line,
arc=3pt,colback=#1,colframe=white,
before upper={\rule[-3pt]{0pt}{10pt}},boxrule=1pt,
boxsep=0pt,left=2pt,right=2pt,top=1pt,bottom=.5pt}

\newcommand{\boxtext}[2]{\boxcolor[#1]{\small \textsf{\textsc{#2}}}}

\newcommand{\ra}[1]{\renewcommand{\arraystretch}{#1}}

\begin{document}

\maketitle

\begin{figure}[t]
    \centering
     \includegraphics[width=\linewidth]{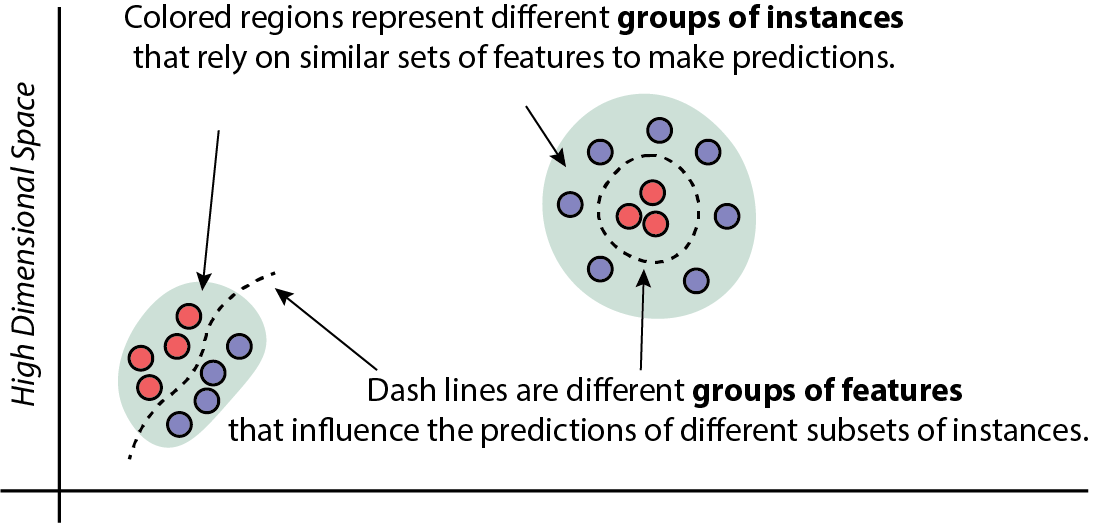}
     \caption{
		 An explanation summary of model explanation is a co-cluster of instances and features based on their similarity of decision attributions.
                }
    \label{fig:illustration}
    \vspace{-3mm}
\end{figure}
\section{Introduction}
``All models are wrong, but some are useful.''
The application of machine learning (ML) models, including deep learning neural networks,
is prevalent in all aspects of human activities 
and nowadays the main driving force of technological advances 
such as self-driving cars, personal assistants and medical diagnoses.
While there are more new models and architectures proposed 
to improve the accuracies of different tasks,
the reason for such popularity also implies that there is no silver bullet for creating the best model.
Hence, the creation of an ML model is a human-centric activity that involves 
lots of reasoning, brainstorming, and most importantly -- understanding processes.
Understanding how a model works on one's own data improves the task performances in a holistic scope 
not only limited to the model design but also the data preprocessing and feature engineering steps.
Yet, ML models nowadays introduce a challenging problem on their \textit{interpretability}.
It becomes so complex to understand what the models have learned 
that using them as a black box could result in 
adversely affecting people's safety, financial, or legal status \cite{voigt2017eu}.

Thus, explainable Artificial Intelligence (XAI) becomes an emerging research field 
where a lot of efforts have been devoted to extracting the logic behind how the models think when making decisions. 
Overall, these logical models focus on the usage of \textit{decision trees}, \textit{rules}, and \textit{instance-level feature importance} 
to mimic or customize the behavior of the ML models \cite{guidotti2018survey} 
so that people can understand how the model works through decision paths or scoring systems. 
In particular, \textit{instance-level feature importance} explanations become more popular to explain sophisticated models. It produces an accurate \textit{local} explanation as they only focus on a single instance. 
Such customization can even allow the explainer to be embedded in the ML model \cite{bien2011prototype,chen2019looks,kim2014bayesian,li2018deep}. 
Thus, local explanations have been readily proposed not only in explaining tabular data classification \cite{lundberg2017unified,ribeiro2016should} but also complex deep learning tasks in natural language processing \cite{poerner2018evaluating} and computer vision \cite{chen2019looks}.

Of course, the ability to customize does not come as a free lunch. 
As the explanation is tailored towards an individual instance, 
the explanation model loses the advantages of providing aggregated explanations to generalize on the whole dataset. 
This limits its usage on providing simple textual information or visualization to describe how the model works. 
Such a limit, however, is where visualization techniques come in handy.
We observe that most feature importance based explanations in current literature can be \textit{summarized} \cite{chandola2007summarization} into an explanation summary. 
The goal of summarization (Figure~\ref{fig:illustration}) is to find a compact description of the dataset with a minimum cost of information loss (i.e., information theoretic). 
In other words, it finds a explanation summary of the ML model, which is the groups of instances with similar explanations(colored regions) and the groups of features that are used to explain similar sets of instances (dashed lines). 
Thus, the summary is a compact \textit{global explanation} that enables effective visualization to communicate a model's general behavior. 
To fill the gaps of local explanation techniques on XAI tasks, 
we propose a scalable data summarization technique that only takes the generic form of the explanation information into account 
so that we can leverage the existing explanation techniques on different domains to provide useful visual data summaries. 
In \systemname, we show that our implementations helps establish a holistic workflow for XAI experience concerning tabular data, texts or images. 
In short, our contributions are as follows:

\begin{itemize}[noitemsep,topsep=0pt,leftmargin=3mm]
	\item \textbf{\algoname, a scalable algorithm that generates a compact data summary for an ML model and input data.}
	It takes any generic \textit{feature importance} based explanations from the model and works for both structured and unstructured data. 
	The algorithm consists of (1) an information-theoretic model to determine the best data summary and 
	(2) an efficient sketching technique to speed up the computation. 
	In Section~\ref{sec:exp} we show that \algoname produces meaningful results and scales to large data.
	
	\item \textbf{\systemname, an interactive system for scalable interpretation and exploration of the input data and the ML model together.} 
	By leveraging our algorithm to group similar instances and explanations, 
	we enable a seamless workflow that connects different needs regarding the global, local, and class explanations in the current XAI systems \cite{liao2020questioning}.
	
	\item \textbf{Three use cases covering ML model interpretations on tabular data, image, and text.} 
	We demonstrate that our algorithm and system enhance the XAI user experiences on model interpretability to three mainstream data analysis. 
 \end{itemize}

\begin{figure*}
    \centering
     \includegraphics[width=\linewidth]{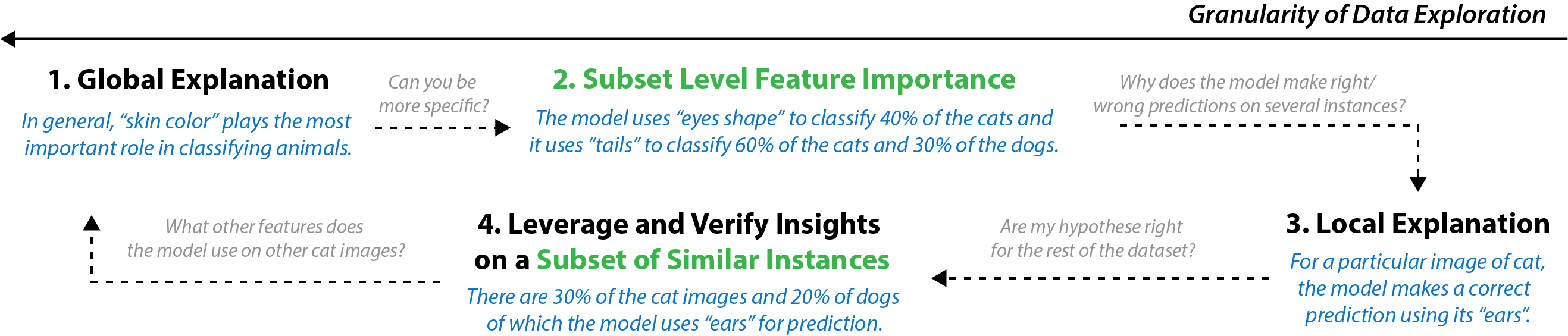}
     \caption{
		 A workflow illustration of how reducing the granularity of global explanation 
		 and increasing the granularity of local explanation comprehend the analysis 
		 on the machine learning model and the dataset. 
		 An \textbf{\textcolor{darkpastelgreen}{explanation summary}} that groups similar features and instances opens the opportunity of addressing different tasks in explainable machine learning.
                }
    \label{fig:need}
\end{figure*}

\section{Related Work}
\label{sec:relwork}
To facilitate human understanding towards complex models through visualization, research mainly focus on visualization on three aspects:
model internals, logics induced from the models, and instance level feature vectors that describe the behavior of the model.

\subsection{Visualization of Model Internals}
\label{sec:relwork_internals}
Visualization has been applied readily to understand and interact with deep learning neural networks. 
In fact, a survey about deep learning visual analytics by Hohman \etal \cite{hohman2018visual} has listed more that 40 representative works in this area in the last 5 years. 
We encourage the readers to read the survey paper for a deeper investigation to the subject.

The simplest form of a neural network can be represented by a node-link diagram in which each node represents a neuron and link presents a connection weight between two neurons \cite{tzeng2005opening}. 
As nowadays the ways neurons are connected become more sophisticated and opaque, 
various visual analytics approaches have been developed to understand different properties of the networks. 
RNNVis \cite{ming2017understanding} and LSTMVis \cite{strobelt2017lstmvis} address the understanding of recurrent neural network (RNN) by visualizing the bipartite relationship between hidden memories and input data and hidden memory dynamics with parallel coordinates respectively. 
Autoencoder is addressed by Seq2SeqVis which proposes a bipartite graph visualization to visualize the attention relationships between input and its possible translations to enable model debugging \cite{strobelt2018s}. 
Another type of popular models for image classification is Convolutional Neural Networks (CNN). 
CNNVis \cite{liu2016towards}, Blocks \cite{bilal2017convolutional}, AEVis \cite{liu2018analyzing} and Summit \cite{hohman2019s} are graph visualizations that 
aggregate similar neurons, connections, and similar activated image patches to convey learned visual representations from the model.

Besides visualizing the structures of a neural network, 
there are visual analytics systems that assist the model development processes in the industry. 
ActiVis \cite{kahng2017cti} is a visual analytics system used by Facebook to explore industrial deep learning models. 
Google has developed TensorFlow Graph \cite{wongsuphasawat2017visualizing} and what-if tool \cite{wexler2019if} to help developers understand and test the behavior of different ML models.

The work in these criteria mainly addresses the visual analysis for model developers who have sufficient knowledge of the methodologies of their models. 
However, a more general AI tool requires the assessment and involvement of end-users, decision-makers, and domain experts. 
Addressing the needs of border XAI user experience, our work focuses on providing general explanations of ML models to users without requiring them to know the architectures.


\subsection{Visualization of Logical Models}
\label{sec:relwork_surrogate}
Logical models like decision trees \cite{craven1996extracting} or rules \cite{martens2008decompositional,yang2017scalable} can address the interpretation of complex models by using them to infer an approximated model from any ML models. 
Given a set of test data, the original model gives the predictions and the logical models use them as labels to train another classifier. 
The resulting classifier can be used to mimic the behavior of the original model while providing good interpretability to the users. 
Through visualizing logical models, users gain knowledge of the model's capability.

Rule Matrix \cite{ming2018rulematrix} is proposed to build and visualize the surrogate rule list to understand the model's behavior by interacting with the rules. 
Gamut \cite{hohman2019gamut} uses generalized additive models to construct a set of linear functions for each feature in the dataset to understand models through line charts. 
TreePOD \cite{muhlbacher2017treepod} and Baobabview \cite{van2011baobabview} visualizes the decision trees with different metrics incorporated for model understanding. 
iForest \cite{zhao2018iforest} visualizes random forests with data flow diagrams to interact with multiple decision paths.

For complex model, using logical models to explain the complex model is the consideration of \textit{fidelity} -- 
the accuracy of the explanation on the original model. 
It creates an additional layer of performance concerns. 
Therefore, local explanation methods are proposed to explore the possibility to provide accurate explanations 
or even be embedded in the original model training process. 
Yet, they only return results for an instance and do not consider a global explanation to the whole dataset, 
our work addresses the challenges of visually constructing a global view for local explanations.

\subsection{Feature Vector Visualization}
\label{sec:relwork_feature}
Local explanation models give feature scores for each instance. The features can be the features from original data \cite{lundberg2017unified,ribeiro2016should,shrikumar2017learning} or a set of external information like concepts \cite{kim2018interpretability} or training data \cite{chen2019looks,li2018deep,ming2019interpretable}. Visual analysis can be directly applied to interact with the features \cite{ming2019protosteer} or the feature vectors can be visualized as a matrix where rows represent the instances and columns represent the features \cite{sawada2019model}. 

Besides, the data comes out from a deep neural network can appear as embeddings such that the linear distances between vectors represent their similarities as the model's rationale. The main visualization technique to understand these feature vectors is projection \cite{grover2016node2vec,li2018embeddingvis,liu2017visual,pezzotti2017deepeyes,rauber2016visualizing,xiang2020interactive}. Treating the embedding as high dimensional data projection techniques such as tSNE, MDS, or PCA are applied to discover semantic groups inside the dataset from the resulting scatterplot. Users can assess the ML model and refine the original data from the brushing the filtering interactions in the projection.

Our technique identifies the scalability and usability challenges in the existing visualization. The projection technique mainly suffers from cluttering and the lack of feature information in the visualization which is crucial for a comprehensive explanation. \algoname aims at providing compact representations of both data and features so that visual information is more precise. Also, we address the needs of explanation exploration with different granularity by the proposed analytic workflow illustrated by \systemname, providing new ways to extend the powerful local explanations to scalable visual analytics.


\section{Tasks Analysis of XAI Systems} 
\label{sec:tasks}
Before we propose our methodology to generate an explanation summary for feature importance based explanations,
we first review the taxonomy of XAI tasks to induce the reasons \textit{how can a visual explanation summary of ML model help}.
By understanding the tasks,
we can consolidate the design considerations to expand our techniques into an effective user interface.
To explain the tasks and the use of data summary systematically, we use a simple workflow of XAI (Figure~\ref{fig:need}) to connect the essential relationships among three main XAI tasks~\cite{liao2020questioning}: \textbf{Global}, \textbf{Local}, and \textbf{Class} explanations.

\noindent\textbf{T.1 Global Explanation.} The goal is to understand the overall weights of features used by the model to explain how AI makes decisions on the dataset in general. For example, imagine we have a model that tells what animal does an image contains. 
To understand the model, the first question a user might ask is \textit{what features the model uses to make a prediction?}
An XAI technique might give us a sorted list of features based on their influences to the model (Figure~\ref{fig:need}.1) -- 
it tells that ``skin color'' is the most important factor.  

\noindent\textbf{T.2 Class Explanation.} 
Understanding how and whether the model works in each class allows users to understand the decision boundaries in a smaller granularity to develop insights.
For example, from a global explanation, ``eyes'' are used to predict many cats. ``Eyes'' and ``cat'' are the key information to understand the model rationale in a local region.

\noindent\textbf{T.3 Local Explanation.} 
For verification and inspections in full details, users need to inspect all explanatory features of a predicted instance (Figure~\ref{fig:need}.3).
For example, why is there a cat predicted as ``dog''? The difference between instances' behavior can evaluate important decision boundaries. We can know that the image's cat has white skin, which is a ``dog'' feature by inspecting a single image. 

\noindent\textbf{Usefulness of Explanation Summary.} First, it can provide a global explanation with a better granularity (Figure~\ref{fig:need}.2). Instead of aggregating the whole dataset to rank the features, it tells directly that ``eyes'' are used on many cats while ``ears'' are used on many dogs. 
This answer avoids the mirage of aggregated features over different subsets.
Also, clustering instances allows users to go from local to global explanation.
For example, by browsing a cat image and knowing it has a wrong prediction due to its white skin, we might want to know all the cat images with the ``white skin'' feature will be predicted as ``dog''.
By inspecting all the cat images or images that have ``white skins'' (Figure~\ref{fig:need}.4), 
we go back to the inspection of a group of images again.

In detail, the tasks in the above workflow generalize the studies that consolidate the key user requirements of model explainability \cite{adadi2018peeking,carvalho2019machine,gilpin2018explaining,guidotti2018survey,mohseni2018survey,ras2018explanation}.
Furthermore, there is a plenty of empirical studies on the requirement of XAI from industry practitioners \cite{amershi2019guidelines,boukhelifa2017data,hohman2019gamut,holstein2019improving,liao2020questioning,muller2019data,rule2018exploration}.
They provide good empirical evidence from real experts to outline the guidelines for designing XAI systems. The details of \textbf{T.1-3} derived from these studies are provided in Appendix~\ref{sec:tasks_appendix}.

\section{Machine Learning Model Explanation Summary}
\label{sec:model}
In this section, we describe the definition of a model explanation summary as well as the algorithms to compute it from the local explanations.

\subsection{Generic Representation of Local Explanation}
\label{sec:data}
The most generic form of a local explanation $e_i$ is a feature vector with $n$ total number of explanatory features used in explaining the whole dataset. 
Each value in the feature vector $e_{ij}$ is the explanation importance of feature $j$ on the instance $i$ (e.g. ``skin color'' on a cat image). 
For more details of local explanation techniques, we redirect readers to Appendix~\ref{sec:background}. 
All $m$ instances' explanations from the whole dataset thus can be expressed as a real-value matrix $E  \in \mathbb{R}^{m,n}$. 
One important property of the matrix is that it is \textit{sparse} i.e. $nm \gg nnz(E)$ where $nnz(E)$ is the number of nonzeros in $E$. 
It ensures the explanation to use a small number of features to explain the model behavior so that the decision logic would not overwhelm the user. 
Also, to simplify the discussion afterwards, 
we assume $e_{ij} \geq 0$ and $\sum_{i,j}e_{ij} = 1$
\footnote{All matrices can satisfy these properties with a min-max scaler (on its absolute values if the sign does not matter).}.

\begin{table}[tb]\centering
    \ra{1.3}
    \begin{tabular}{@{}llcl@{}}\toprule
        &\multicolumn{1}{c}{Data Type} & \phantom{abc} & \multicolumn{1}{c}{Explanatory Features} \\ \midrule 
        \\[-3mm]
        &Tabular & &  \makecell{A set of \textbf{logics} (ranges). \\ e.g. It \underline{will} rain since \textcolor{amber(sae/ece)}{$\textit{precipitation} \geq 90\%$}.} \\
        \\[-3mm]
        &Image & & \makecell{A set of \textbf{common visual representations}. \\ e.g. It is a \underline{pig \raisebox{-.3\height}{\coloremoji{🐷}}} since I notice its \textcolor{amber(sae/ece)}{\textit{nose}} \raisebox{-.3\height}{\coloremoji{🐽}}.} \\ 
        \\[-3mm]
        &Text & & \makecell{A set of \textbf{topics}. \\ e.g. This review is \underline{positive} since there are \\ words like \textcolor{amber(sae/ece)}{\textit{excellent / fantastic / amazing}.}} \\ 
    \end{tabular}
    \caption{Intrinsic nature of \textcolor{amber(sae/ece)}{explanations} among tabular, image and text data. It affects how we construct the explanation matrices.}
    \label{table:nature}
\end{table}

\subsection{Explaining Tabular, Image, and Text Instances}
Given a generic form of instance explanation, we now drill down to an in-depth discussion of how these feature vectors can be applied to explain tabular, image, and text instances. Although all of them result in explanation matrices, their intrinsic nature, shown in Table~\ref{table:nature}, affects how we modify our modeling pipelines to construct the features (i.e., columns)  to acquire meaningful explanations. We provide end-to-end data modeling examples in Appendix~\ref{sec:pipeline}.  

\noindent\textbf{Tabular Data}. Logical models like decision trees and random forests discretize the attributes in the dataset to a set of \textit{logics}. Similarly, the explanatory features should be not only the original attributes but also the different ranges for better diversity. For example, all cities (instances) rain based on their precipitations (attribute), but how each city rain in different percentages of precipitation (ranges) reveals different climates. \looseness=-1

\noindent\textbf{Image Data}. Users normally classify images with the common visual features among the same entity (e.g., stripes in zebras). Similarly, an image in an ML model can be explained with representative image patches collected from the original data that unify the reasoning process with a limited set of features instead of pixels in each image.

\noindent\textbf{Text Data}. Multiple documents are usually explained with common topics instead of single phrases because similarly meant data can be totally different words. For example, ``good'' and ``great'' represent similar sentiments. Thus, the explanatory features should be a set of topics instead of words to avoid an overly sparse matrix.

\subsection{Problem Definition}
\label{sec:def}
The information-theoretic goal for summarizing the explanation matrix is to group similar instances and explanatory features simultaneously that balance compactness and information loss.
Let $R$ and $C$ be the set of row (instance) and column (feature) vectors in $E$ respectively such that $E$ is equivalent to a joint distribution between $R$ and $C$ (i.e. $p(R,C)$).
Our goal is to find the optimal row and column clusters $\hat{R}$ and $\hat{C}$ so that it presents the explanation summary in Figure~\ref{fig:illustration}.
Therefore, the first question is, how we should measure the information loss?
For example, consider the following synthetic explanation matrix below:

\begin{equation*} 
p(R,C) = 
\begin{bmatrix}
.1 & .1  & 0 & 0\\ 
.1 & .1  & 0 & 0\\ 
0 & 0  & .2 & .2\\ 
0 & 0 &  0 & .2
\end{bmatrix}
\end{equation*}
It is obvious to group the rows into two clusters: $\hat{r}_1 = \{r_1,r_2\}$, $ \hat{r}_2 = \{r_3,r_4\}$
and the columns into two clusters: $\hat{c}_1 = \{c_1,c_2\}$, $\hat{c}_2 = \{c_3,c_4\}$.
The information theoretic definition of the resulting compression $p(\hat{R},\hat{C})$ 
and the approximation matrix recovered from the compression $q(\hat{R},\hat{C})$ are as follows \cite{dhillon2003information}:
\begin{equation*}
p(\hat{R},\hat{C}) = 
\begin{bmatrix}
.4 & 0 \\ 
0 & .6
\end{bmatrix}
\textit{, }
q(\hat{R},\hat{C}) =
\begin{bmatrix}
.1 & .1  & 0 & 0\\ 
.1 & .1  & 0 & 0\\ 
0 & 0  & .133 & .267\\ 
0 & 0 &  .067 & .133
\end{bmatrix}
\end{equation*}
Each entry in the approximation matrix $q(\hat{R},\hat{C})$ is calculated as follows:
\begin{equation}
    q(r,c) = p(\hat{r},\hat{c}) \times \frac{p_{R}(r)}{p_{\hat{R}}(\hat{r})} \times \frac{p_{C}(c)}{p_{\hat{C}}(\hat{c})}
\end{equation}
For example, $q(3,4) = .6 \times (.4)/(.6) \times (.4)/(.6) = 0.267$.
Thus, the compression loss can be expressed with  metrics such as Kullback-Leibler (KL) divergence of $p(R,C)$ from $q(\hat{R},\hat{C})$: 
\begin{equation}
D_{KL}(P,Q)=\sum_{x \in \chi }P(x)log(\frac{P(x)}{Q(x)})
\label{eqn:kl}
\end{equation}
Yet, we observe a shortcoming of directly using KL divergence on the whole matrix. The $P(x)$ in Equation~\ref{eqn:kl} tells us that each entry's contribution to the result is not independent to the clusters that it does not belong to. Therefore, we propose a loss function $D(\hat{R},\hat{C})$ such that each entry's loss is marginal to its row and column cluster:
\begin{equation}
    \begin{aligned}
    D(\hat{R},\hat{C}) = \sum_{\hat{r} \in \hat{R}}D_{KL}(P(r \in \hat{r}, C),Q(r \in \hat{r}, C)) \\
    +\sum_{\hat{c} \in \hat{C}}D_{KL}(P(R, c \in \hat{c}),Q(R, c \in \hat{c}))
    \label{eqn:loss}
    \end{aligned}
\end{equation}
Such a marginalization prevents entries with high values dominating the calculation result, which we will demonstrate the subsequent improvement in Section~\ref{sec:exp}.

Once we quantify the information loss, the next challenge is how should we choose the number of row and column clusters?
If we do not cluster any rows and columns at all, $D$ will equal to zero.
Whereas if we only have one cluster, the loss will be huge.
Yet, neither of them is a summary of the data as it either represent the original matrix or a summary with poor quality.
To automatically determine the optimal partitions,
the idea is to use Minimum Description Length Principle (MDL), which states that the best model is the one that minimizes the total description length of the expression: $model$ (i.e., number of clusters $\left \| \hat{R} \right \|$ and $\left \| \hat{C} \right \|$) $-$ $correction$ (i.e., information loss $D$).
Putting them all together, we can now write the total cost function $T$ as:
\begin{equation}
    T(\hat{R};\hat{C}) = \beta_R  \left \| \hat{R} \right \|   + \beta_C \left \| \hat{C} \right \|  +  D(\hat{R},\hat{C})
    \label{eqn:objective}
\end{equation}
which we try to minimize it with the best rows and columns partitions.
$\beta_R$ and $\beta_C$ are user defined parameters to penalize large number of clusters.
Users can increase the values to produce fewer clusters.

\subsection{The \algoname Algorithm}
We now present our \algoname (\textbf{\underline{M}}achin\textbf{\underline{E}} \textbf{\underline{L}}earning M\textbf{\underline{O}}\textbf{\underline{D}}el Summar\textbf{\underline{Y}}) algorithm.
In the previous section we have created our goal 
to find the row and column clusters that minimize the cost function in Equation~\ref{eqn:objective}
among all possible number of clusters and all possible rows and columns combinations.
Yet, the equation itself does not tell us how to reach the solution efficiently.
Since the matrix can be considered as a graph where each entry is a weighted edge between a row node and a column node,
we can use graph summarization \cite{navlakha2008graph} approach to provide a baseline solution (Algorithm~\ref{algo:coco}).
The overall idea is as follows:
\begin{enumerate}
    \item Each row and column starts in its own cluster. Then, we put the row and column clusters into two separate lists (line 1-2).
    \item We first fix the column cluster assignment. For the row clusters in the list, we randomly select a row cluster (line 5).
    \item We compare the selected row cluster with the remaining row clusters in the list as merge candidates (line 7-12):
          we try merging the selected cluster with each remained cluster and calculate the cost reduction by Equation~\ref{eqn:objective} (line 8).
          We choose the candidate that produces the least cost.
    \item If merging the selected cluster and its best candidate reduces the total cost, then we merge two clusters in the list (line 14-15). 
          Otherwise, we remove the selected cluster in the list (line 17). Either way, the list will have one fewer item. 
    \item We repeat steps 2-4, but we fix the row clusters and merge the column clusters instead. 
          The whole algorithm stops until there are no clusters remained in both lists.
\end{enumerate}

\begin{algorithm}

\SetKwFunction{cumprod}{cumprod}
\SetKwFunction{length}{length}
\SetKwFunction{zeros}{zeros}
\SetKwFunction{ceil}{ceil}

\SetKwInOut{Input}{Input}
\SetKwInOut{Output}{Output}

\caption{\algoname (\textbf{\underline{M}}achin\textbf{\underline{E}} \textbf{\underline{L}}earning M\textbf{\underline{O}}\textbf{\underline{D}}el Summar\textbf{\underline{Y}})}
\label{algo:coco}
\Input{%
        \xvbox{5mm}{$R,C$} -- instances and explanatory features\\
        \xvbox{8mm}{$\beta_R$, $\beta_C$} -- regularization terms
        }
\Output{%
        \xvbox{5mm}{$\hat{R},\hat{C}$} -- row and column clusters
        }

    \BlankLine 
    
    \xvbox{2mm}{$\xvar{R}$} $\leftarrow$ $[\{r_1\},\{r_2\},...,\{r_m\}]$, \xvbox{2mm}{$\hat{R}$} $\leftarrow$ $\{\}$ \tcc*{intialize rows}

    \xvbox{2mm}{$\xvar{C}$} $\leftarrow$ $[\{c_1\},\{c_2\},...,\{c_n\}]$, \xvbox{2mm}{$\hat{C}$} $\leftarrow$ $\{\}$ \tcc*{intialize columns}

    \xvbox{5mm}{$loss$} $\leftarrow$ $D($\xvar{R}$,$\xvar{C}$)$ \tcc*{initialize loss function}

    \While{ $size(\xvar{R}) > 0$ and  $size(\xvar{C}) > 0$}{

        \xvbox{2mm}{$\xvar{r}_0$} $\leftarrow$ $random\_pop(\xvar{R})$ \tcc*{randomly extract a cluster}
        \xvbox{7mm}{$\Delta L_{max}$}  $\leftarrow$ 0, \xvbox{5mm}{$r_{max}$} $\leftarrow$ \textit{undefined}

        \For{\xvbox{2mm}{$\xvar{r}$} $\leftarrow$ $\xvar{R}$}{
            \xvbox{4mm}{$\Delta L$} $\leftarrow$ $\beta_R - D_{KL}( \{\xvar{r} \cup \xvar{r}_0\},\xvar{C}\cup\hat{C})$

            \If{$\Delta L > \Delta L_{max}$}{
                \xvbox{8mm}{$\Delta L_{max}$} $\leftarrow$ $\Delta L$,
                \xvbox{5mm}{$r_{max}$} $\leftarrow$ $\xvar{r}$
            }
            
        }
        \eIf{$\Delta L_{max} > 0$}{
            \xvbox{5mm}{$r_{max}$} $\leftarrow$ $\{r_{max} \cup \xvar{r}_0\}$ \tcc*{merge two clusters}

        }{
            \xvbox{2mm}{$\hat{R}$}.push($r_0$) \tcc*{push the cluster to final result}
        }

        \tcc{same procedure as for C...}
    }
\end{algorithm}

Overall, in every iteration, a row (column) needs to measure the cost reductions with the remaining candidates in the list, which has the maximum size of $||R||$ ($||C||$).
Therefore, the time complexity of the basic algorithm is $O(||R||^{2} + ||C||^{2})$.
As a quadratic algorithm is infeasible for any moderately sized data for exploratory visual analysis,
we now propose a speed-up strategy to make our algorithm suitable for interactive performance.
\subsubsection{Speed Up Strategies With Data Sketches}

While a randomized bottom-up algorithm scales linearly, 
Algorithm~\ref{algo:coco} is time-consuming as it needs an extra loop to compare all possible row or column clusters (line 7) in every iteration.
However, if we look at the example matrix in Section~\ref{sec:def},
it is obvious that the first two rows (columns) are completely different from the last two rows (columns).
Comparing candidates that are different indeed is of no use since they are unlikely to reduce the total cost.
Thus, to speed up the algorithm, 
we propose a k-nearest neighbor query strategy with a novel use of locality sensitive hashing (LSH) \cite{charikar2002similarity} scheme to encode a row of column clusters.
LSH defines a family of hash functions (i.e., sketches) $[h_{1}(v_i),h_{2}(v_i),...,h_{n}(v_i)]$ for a vector $v_i$
so that the probability of hash collisions between two vectors is proportional to their euclidean distances (i.e., $sim(v_i,v_j)\sim Pr[h_k(v_i)=h_k(v_j)]$).
Vectors with similar values thus can be stored in the same buckets in an LSH table.
Furthermore, we can extend this proportional to retrieve similar row (column) clusters.
If two clusters have many similar vectors, then the number of hash collisions will be high.
Therefore, the top-k clusters from the query will likely be similar neighbors.

The query algorithm is illustrated in Algorithm~\ref{algo:lsh}.
First, an LSH table needs to be built for rows and columns, respectively.
Then, when a neighbor query is performed, 
we can use the hash keys from the query's vectors to perform a table look-up to retrieve all the collided entries with the entries in the cluster ( subroutine $query\_lsh\_table$ in line 2).
We count the average number of collisions between the entries from the query cluster and the ones from the candidate clusters (line 5) and return the top k clusters with the highest number.
This can drastically reduce the number of comparisons and the running time when the matrices are large (Section~\ref{sec:exp}).

\subsubsection{Strategies Addressing Skewness and Sparsity}
\label{sec:heuristics}
Empirically, we observe two challenges when computing the results from real datasets in which we provide the following heuristics to address the problems and demonstrate the effectiveness in Section~\ref{sec:exp}:

\noindent\textbf{Smoothing the explanation values}: When an explanation model assigns values to important features of an instance, the values can be very high (e.g., extremely sensitive features). It could affect the calculation of loss function (Equation~\ref{eqn:loss}) and prevent instances with similarly activated features from being grouped. Therefore, to have an even data distribution in the explanation matrix, we set the maximum value to the knee point of the overall value distributions in the matrix using a knee finding algorithm\cite{satopaa2011finding}. 

\noindent\textbf{Pre-clustering for a cold start in a sparse environment}: Given a sparse explanation matrix, the bottom-up approach might face difficulties in cluster entries when the cost function is stuck a local minimum. Also, as the matrix is sparse, it is hard for the algorithm to know whether there are cluster structures at the beginning. These adversely affect the formation of significant clusters. To address this cold start problem, we reference from spectral graph partitioning \cite{dhillon2001co} to create relatively smaller partitions of rows and columns using their singular vectors from SVD decomposition. Then, we can use our information-theoretic objective function to compress the matrix further.

\begin{algorithm}[tb]
    \label{algo:lsh}
    \SetKwFunction{cumprod}{cumprod}
    \SetKwFunction{length}{length}
    \SetKwFunction{zeros}{zeros}
    \SetKwFunction{ceil}{ceil}
    
    \SetKwInOut{Input}{Input}
    \SetKwInOut{Output}{Output}
    
    \caption{Top-k Nearest Neighbor Query}
    \label{algo:lsh}
    \tcc{Initialize \xvbox{3mm}{$\xvar{T}_R$} $\leftarrow$ $build\_lsh\_table(\xvar{R})$ and 
    \xvbox{3mm}{$\xvar{T}_C$} $\leftarrow$ $build\_lsh\_table(\xvar{C})$ after line 2 in Algo.~\ref{algo:coco}}

    \tcc{Replace $\xvar{R}$ with $query(\xvar{r},\xvar{R},\xvar{T}_R,k)$ in line 7 of Algo.~\ref{algo:coco}}
    \Input{%
            \xvbox{2mm}{$\xvar{v}$} -- query cluster\\
            \xvbox{6mm}{$\xvar{V}$, $\xvar{T}_v$} -- remaining clusters and LSH table\\
            \xvbox{2mm}{$k$} -- number of neighbors
            }
    \Output{%
            \xvbox{4mm}{$\xvar{knn}$} -- top k nearest neighbors
            }
    \xvbox{9mm}{$counter$} $\leftarrow$ $Counter()$ \tcc*{initialize counter}
    \xvbox{12mm}{$neighbors$} $\leftarrow$ $query\_lsh\_table(\xvar{v},\xvar{T}_v)$ \tcc*{get collided entries}

    \For{\xvbox{2mm}{$n$} $\leftarrow$ $neighbors$}{
        \For{$\bar{\xvar{v}}$ in $\xvar{V}$}{
            \If(\tcc*[h]{collision between the clusters}){$n$ in $\bar{\xvar{v}}$ } 
            {  
                \xvbox{12mm}{$counter[\bar{\xvar{v}}]$} $+= 1 / |\bar{\xvar{v}}|$ \\
                break
            }
        }
    }   
    \xvbox{4mm}{$knn$} $\leftarrow$ $counter.most\_common(k)$   
\end{algorithm}
\begin{figure*}
    \centering
     \includegraphics[width=\linewidth]{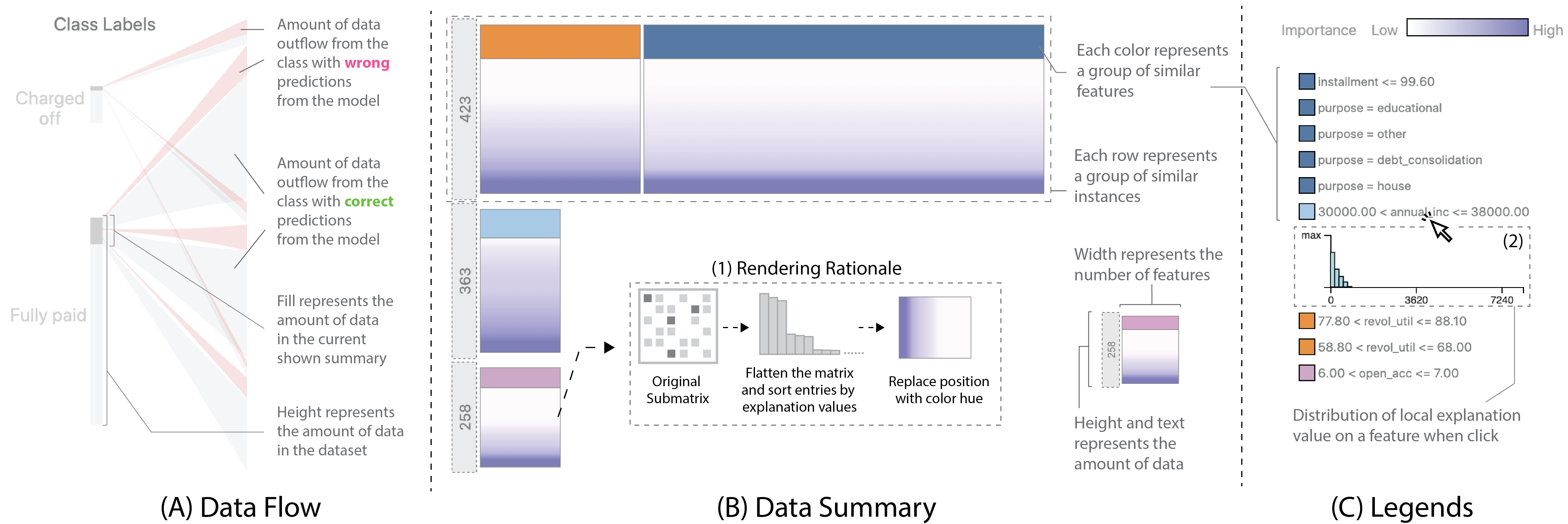}
     \caption{
        Explanation Summary Visualization Design. A: The data flow visualizes the data flow from each class to different instances clusters, providing a sense of overall data distribution among different model decisions.
        B: The summary adjacency list presents each instance cluster as a row, and each explanatory feature cluster as a color.
        The explanation values inside each co-cluster are visualized as a gradient generated in (1).
        C: The legends show the mappings between the features and the color encodings of their respective feature clusters. 
                }
    \label{fig:summary_vis}
\end{figure*}

\section{Design Considerations}
\label{sec:requirement}

Based on Section~\ref{sec:tasks} and Section~\ref{sec:model}, we distill the main design considerations for an interactive visualization interface for addressing a holistic XAI workflow and the summary's characteristics. Considerations in \boxtext{lavenderblush}{pink} address the tasks in Section~\ref{sec:tasks} and those in \boxtext{lavender(web)}{blue} address the data perspective in Section~\ref{sec:model}.  

\begin{itemize}[noitemsep]
    \item[\textbf{C.1}]\boxtext{lavenderblush}{Visual Summary} \textbf{Synthesize instance and feature summary.} 
    Clusters of instances and clusters of features should be displayed together to understand the decision boundaries(\textit{knowledge of a model}) on different subsets (\textit{the influenced population}).

    \item[\textbf{C.2}]\boxtext{lavender(web)}{Sparse Summary} \textbf{Scalable visualization for large sparse data.} As the local explanations are highly customized and independent, the explanation summary will also be a sparse matrix. The visualization needs to highlight small but significant co-clusters.

    \item[\textbf{C.3}]\boxtext{lavenderblush}{Prediction Outcome} \textbf{Display instances' outcomes from the model.} Knowing when and where the model can fail is essential to understand its capability. Thus, the prediction outcome should be embedded in the visual summary and explanations.

    \item[\textbf{C.4}]\boxtext{lavenderblush}{Filtering} \textbf{Filtering data summary by classes or features.}
    When a user collects insights from local or class explanations, the insights need to be verified on a larger population. Thus, filtering by classes or features act as a query from a local analysis to refine the explanation summary in a global view. 
    \item[\textbf{C.5}]\boxtext{lavender(web)}{Level-of-detail} \textbf{Different level-of-detail presentations for tabular, image and text data.} Level-of-details may come in different forms for different data primitives. Although the explanation summaries are the same, when users drill down on details, the presentations should be different.

    \item[\textbf{C.6}]\boxtext{lavenderblush}{Explanation in a Loop} \textbf{Connecting local, global, and class explanations as a loop.} The main three themes of XAI should be connected for a complete ML model explanation (Figure~\ref{fig:need}). Different views related to different scopes of explanations should be tightly integrated.
\end{itemize}

\section{\systemname}
\label{sec:system}
Based on our design considerations in Section~\ref{sec:requirement} and our \algoname algorithm in Section~\ref{sec:model}, we present \systemname, an interactive system for helping users to understand an ML model's decisions on an input dataset \footnote{The system can be accessed at \url{http://128.238.182.13:5004/}}. The interface consists of (A) a main explanation summary visualization, (B) an original data subset view from a selected summary, and (C) an instance view. In the following discussion, we will focus on the main summary visualization and how an XAI workflow in Figure~\ref{fig:need} is established in a visual analytics fashion. 

\subsection{Explanation Summary Visualization Design}
The explanation summary visualization (Figure~\ref{fig:summary_vis}) contains three visual components: the \textit{data flow}, the \textit{adjacency list}, and the \textit{legends}. The dataflow shows how instances from different classes flow to different instance clusters through a Sankey diagram. The adjacency list displays the data summary from the local explanations. The legends display the features and their corresponding color encodings in the adjacency list.

\subsubsection{Adjacency List}
The main visual component of \systemname is the adjacency list of the explanation summary. The explanation summary is a matrix of two sets: instance clusters and feature clusters. 
The intersection between an instance cluster and a feature cluster is a real-valued submatrix of original explanations. Therefore, the simplest way to present the explanation summary is to directly show the original explanation matrix with rows and columns ordered according to their cluster memberships. However, we found the co-clusters hard to be observed when the matrix is sparse (\textbf{C.2}). Since an ML model's decisions are usually diverse on different subsets of the input data, the clustering will also result in many different row and column clusters. Thus, it becomes difficult for users to notice small clusters. Also, we found the information obtained from the matrix hard to memorize when users perform multiple visual inspections and interactions on different widgets at the same time. For example, when a feature cluster is selected, users inspect the features inside in a separate view. After the inspection, it becomes difficult to recall which feature cluster they have selected inside the matrix. These problems related to sparsity and stimulus have also been identified and thoroughly studied in previous literature \cite{ghoniem2005readability,hlawatsch2014visual,okoe2018node}. 

To explore relevant instances and features in a large sparse matrix (\textbf{C.1-2}), we design an adjacency list visualization (inspired by \cite{hlawatsch2014visual}) to present the explanation summary (Figure~\ref{fig:summary_vis}B). Each row in the adjacency list represents an instance cluster, and each color texture represents a feature cluster. The size of an instance cluster is encoded with text and height. For a feature cluster, the size is encoded with width. Thus, each intersection between the instance and feature cluster forms a block (i.e. a cell in $p(\hat{R},\hat{C})$). The blocks in each row are sorted by their values in $p(\hat{R},\hat{C})$. In this arrangement, we fix the instances' positions for users to locate a subset of data easily. Also, the features are color encoded so that users can reference an explanatory feature easily by its color, which helps navigate the features across different widgets (\textbf{C.6}). Furthermore, as the column position restriction is removed, the adjacency list becomes more compact. We acknowledge that categorical color scheme might impose a scalability issue, thus we combine the colors with textures \raisebox{-1ex}{\includegraphics[scale=0.15]{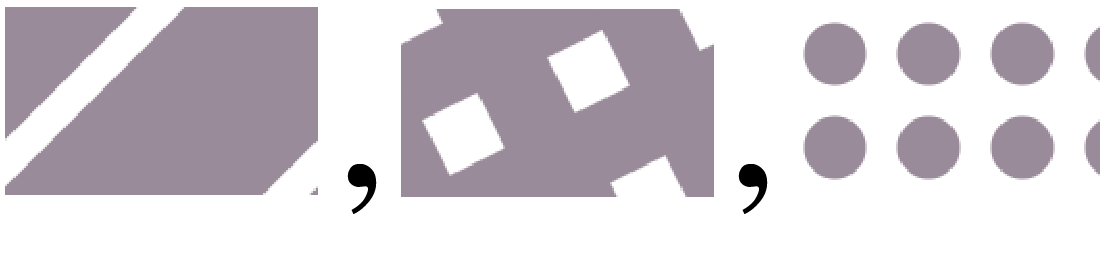}} to increase the available selections.

\noindent \textbf{Visualizing Local Explanation Values} Each block is a co-cluster between a group of instances and features. Thus it is also a sub-matrix of the original explanation matrix. As a sub-matrix contains a distribution of positive real numbers, we display such information as a histogram encoded by a color gradient (Figure~\ref{fig:summary_vis}\clabel{1}). The values in the sub-matrix are sorted from high to low and then encoded by a sequential color scheme. The sorting can provide better clarity on the quality of co-clusters under the sparse matrix clustering condition (\textbf{C.2}).  

\begin{figure*}
    \centering
     \includegraphics[width=\linewidth]{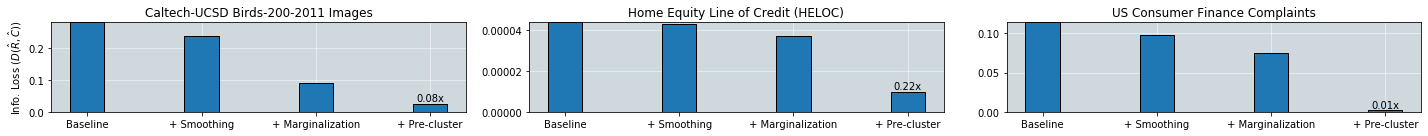}
     \caption{
       Information loss of three datasets' explanation summaries after applying the heuristics from left to right (with final loss reduction shown).
                }
    \label{fig:exp_quality}
\end{figure*}

\subsubsection{Data Flow}
To provide a picture of how data and predictions are arranged in the summary, a Sankey diagram is displayed (Figure\ref{fig:summary_vis}A) on the left of the adjacency list. A vertical rectangle is shown for each class with height encoded as the number of instances in the dataset. The amount of fill of each rectangle is proportional to the number of instances in the currently shown summary.
The horizontal flows represent the portion of data falling into a designated instance cluster. Different colors in the flows represent the amount of data that is either correctly predicted (grey) or incorrectly predicted (red). It helps users assess the capability of the ML model: its performance on each class and the accuracy of each different decision boundaries (\textbf{C.3}).

\subsubsection{Legends}
The legends (Figure~\ref{fig:summary_vis}C) show the color and texture encodings of the clusters of explanatory features.
The features are sorted based on their existence in the current summary.
When we click on a feature, its distribution of explanation values in the dataset is shown as a histogram (Figure~\ref{fig:summary_vis}\clabel{2}) to allow the inspection of its global importance (\textbf{C.1}).

\subsubsection{Interactions}
The explanation summary can be filtered through various mechanisms. Besides explicitly selecting classes and explanatory features for filtering in the dropdown menus, the statistical properties such as the size of clusters and the average explanation values of a co-cluster allow the summary to be filtered through \textbf{sliding} different thresholds. Also, when clusters are selected, the values in the subset are shown in a parallel coordinates to export the important instances from a sparse cluster (\textbf{C.2}) to the subset view through \textbf{brushing} the axes.

\subsection{Visual Analytics Workflow of ML Model Explanation}
We now describe how to leverage the explanation summary to complete a visual analytics workflow. The interactions between different explanations in Figure~\ref{fig:need} are consistent with \systemname's views.
While the adjacency list acts as a global overview of the ML model explanation, the components in the list can be selected and exported to a more focused class and instance inspection. In return, the adjacency list can use the findings from local explanations for verification or further insights. Thus, the workflow in the system forms a finite state transition among global, subset (class), and instance explanations. 
The discussion below mainly focuses on how the system helps circulate different XAI tasks.

\subsubsection{Global $\longrightarrow$ Subset (Class) Explanation}
After exploring the adjacency list, users can proceed to a subset of the clusters by clicking on a row cluster or a co-cluster(\textbf{zoom and filter}). 
After selecting an explanation subset and extracting the instances with significant values, 
users can proceed to understand the local decision logics from the behavior of instances inside. To provide contextual explanations for tabular, image, and text data, we propose three different ways to visualize the subsets (\textbf{C.5}).

\begin{figure*}
    \centering
     \includegraphics[width=\linewidth]{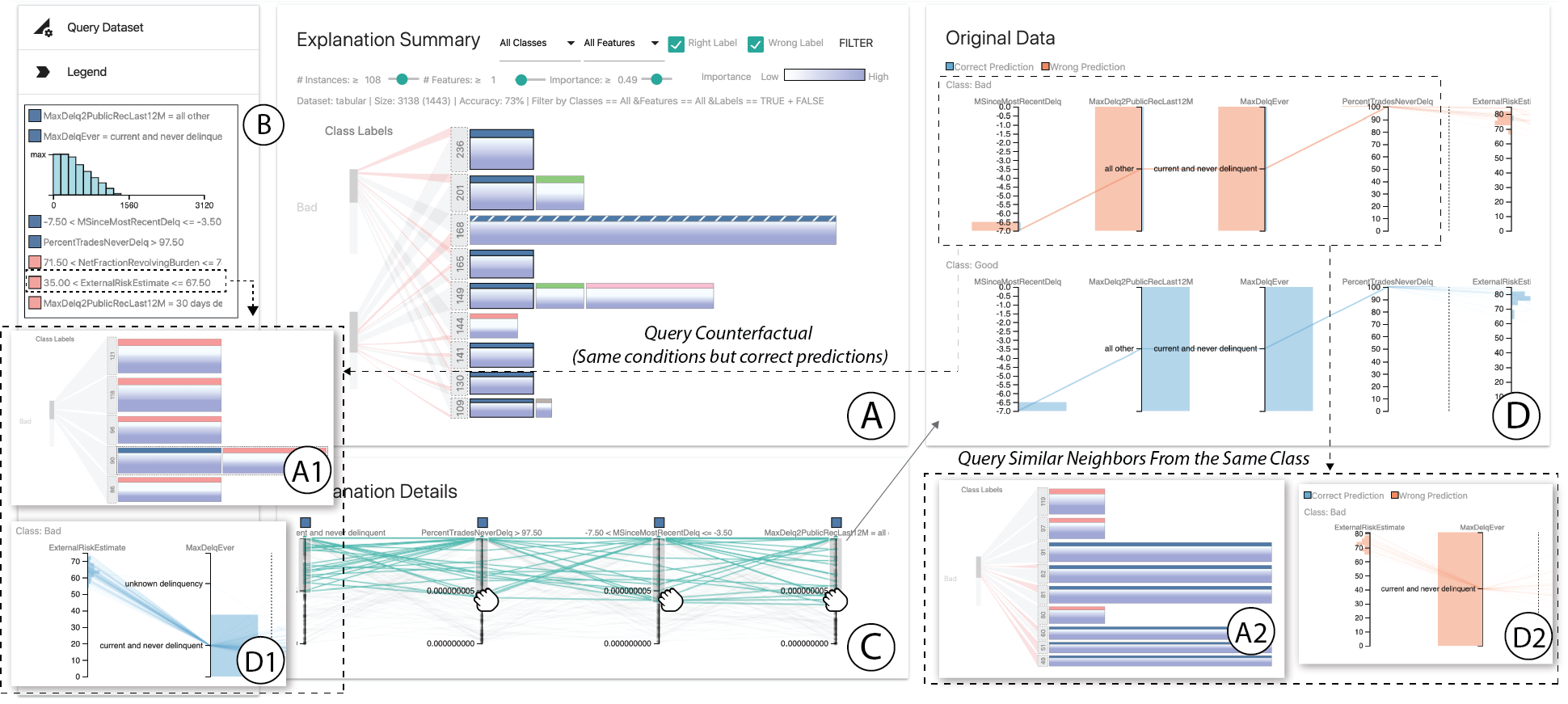}
     \caption{
        Use case of understanding a neural network of credit risk classification trained on tabular data. A, A1, A2: The explanation summary of the whole data, counterfactual of the query, and similar neighbors from the query, respectively.  B: Explanatory features with value distributions to understand the popularity among the dataset. C: Explanation details for filtering and zooming significant explanations. D, D1, D2: Subset views of the selected subsets from the summary.
                }
    \label{fig:case_tabular}
\end{figure*}

\noindent \textbf{Tabular.}
The system visualizes the tabular data in multiple sets of parallel coordinates (Figure~\ref{fig:case_tabular}\textbf{D}). Each set of parallel coordinates represents one class, and the lines inside represent the instances. The axes show the features in the original dataset, and the selected features are positioned at the front. The lines are colored based on whether their predictions. There are also two histograms on each axis that represent the distributions of the correctly and incorrectly predicted instances. 

\noindent \textbf{Image.}
For image data, the system shows the similarly explained instance on one column and their corresponding common visual representations on another column (Figure~\ref{fig:teaser}\textbf{B}). The instances shown are also grouped by their classes and are surrounded by colored frames that indicate the predictions. All instances' and features' images are displayed to acquire a visual impression of similar images and explanations. 

\noindent \textbf{Text.}
The system shows the number of selected instances as bar charts grouped by class and prediction outcome on the left column, and the topics and words that are used to explain the instances on the right column (Figure~\ref{fig:case_text}\textbf{C}). Users can understand what kinds of words are combined to make decisions on each prediction and further select individual words inside each topic to filter the bar charts. When a bar is clicked, the documents can be exported to the local explanation view.

\subsubsection{Subset (Class) $\longrightarrow$ Local Explanation}
After a subset of instances and features are inspected, users can drill down to inspect an instance with full details for insights or hypotheses (\textbf{detail on demand}). Similarly, different arrangements are provided to inspect instances from tabular, image, and text data (\textbf{C.5}).

\noindent \textbf{Tabular Instances.}
The instances are selected by brushing the parallel coordinates in the subset view and rendered in the data table with original features to browse the exact numerical and categorical values.  The color of each cell represents the prediction outcome.

\noindent \textbf{Single Image.}
An image and its top influencing features (image patches with the highest similarity scores) are displayed. The instance and features also have their bounding box of neuron activations to inspect the relationship between different patches.

\noindent \textbf{Text Documents.}
The full documents selected from the bar charts are shown. The words that are explanatory features in the document are highlighted by a sequential color map with their explanation values.  

\subsubsection{Local $\longrightarrow$ Global Explanation}
Users might formulate insights and hypotheses throughout the top-down inspection. For testing the hypotheses, the explanatory features and instances in the local explanation panel are clicked, and their values become the conditions for filtering the adjacency list (\textbf{Query}) (\textbf{C.4}). For tabular data, when a cell in the data table is clicked, the logic that includes the cell's value will be included (e.g., when a categorical cell valued ``\textit{education}'' in column ``\textit{purpose}'' is clicked, the explanatory feature ``\textit{purpose = education}'' will be selected). For image and text data, the user clicks on the image or document for class queries and the image patches or words for feature queries. Overall, users can filter the explanation summary by class, prediction outcome, and explanatory features. As a result, a new and refined overview summary is available to perform global explanation tasks again, which completes the loop.


\section{Evaluation}
To evaluate the scalability and the quality of \algoname, we perform quantitative experiments and use case scenarios on a variety of datasets. 

\subsection{Experimental Setup}
The implementations are written in NumPy, and the experiments are run in a MacBook Pro with 2.4 GHz 8-Core Intel Core i9 CPUs and 32GB RAM. We use the following real-world datasets and ML models to conduct our experiments and use cases:

\noindent\textbf{Caltech-UCSD Birds-200-2011 Images.} The dataset includes 11,788 images with 200 species of birds. We use a Convolutional Neural Network (CNN) with a prototype layer~\cite{chen2019looks} and achieves the highest test accuracy of $73.63\%$. The explanation matrix is extracted from the prototype layer, which has 1330 visual explanatory features.

\noindent\textbf{Home Equity Line of Credit (HELOC).} It contains binary classifications of risk performance (i.e., good or bad) on 10,459 samples with even class distributions. We train a two-layer neural network and achieves the highest test accuracy of $72.59\%$. We extract 167 logics and use SHAP~\cite{lundberg2017unified} to construct our explanation matrix.

\noindent\textbf{US Consumer Finance Complaints.} The dataset contains 22,200 documents with ten classes (e.g., debt, credit card, and mortgage). We train an LSTM neural network model and achieve the highest test accuracy of $84.54\%$. We use IntGrad~\cite{sundararajan2017axiomatic} to generate explanations for words in each document. We further combine the words by clustering their embeddings to generate 500 topics as the explanation features.

\begin{table}[tb]\centering
    \small
    \ra{1}
    \begin{tabular}{@{}llcccc@{}}\toprule
        &\multicolumn{1}{c}{Algorithm} & \phantom{abc} & \multicolumn{1}{c}{Tabular} & \multicolumn{1}{c}{Image} & \multicolumn{1}{c}{Text}\\ \midrule 
        \\[-3mm]
        &Baseline & &  33 mins  & 21 mins & $>7$ hours\\
        \\[-3mm]
        &Baseline + LSH & & \textbf{5s}  & \textbf{13s} & \textbf{9s}\\ \bottomrule
        \\[-3mm]
    \end{tabular}
    \caption{Run time on different datasets.}
    \vspace{-3mm}
    \label{table:exp_runtime}
\end{table}

\subsection{Quantitative Evaluation}
\label{sec:exp}
\noindent\textbf{End-to-end quality evaluation.} To evaluate how each of our heuristics improves the quality of the summarization results, we report the quality (information loss) of the baseline implementation (i.e., straightforward minimization of Equation~\ref{eqn:kl}) as well as the effects of applying marginalization (Equation~\ref{eqn:loss}), smoothing, and pre-clustering from Section~\ref{sec:heuristics}. Overall, the heuristics significantly improve the quality of the result (Figure~\ref{fig:exp_quality}). The final reductions of information loss range from 78\% to 99\%. To visually understand the quality of the summarization results, we provide visual outcomes of the explanation summaries in Figure~\ref{fig:tabular_matrix}-\ref{fig:text_matrix} in the Appendix. \looseness=-1

\noindent\textbf{Effect of data sketches on run time performance.} We report the effect of the run time on the three datasets with the speedup strategies (Algorithm~\ref{algo:lsh}) in Table~\ref{table:exp_runtime}. The result clearly shows that by replacing the quadratic computation in the baseline approach (Algorithm~\ref{algo:coco}), it becomes possible to produce results in interactive time. We also observe that the calculation of information loss is not linear in runtime since there are lots of data slicing operations to compute the approximation matrix ($q(\hat{R},\hat{C})$). The results highlight the importance of limiting the number of candidate comparisons in the bottom-up process. \looseness=-1

\begin{figure*}
    \centering
     \includegraphics[width=\linewidth]{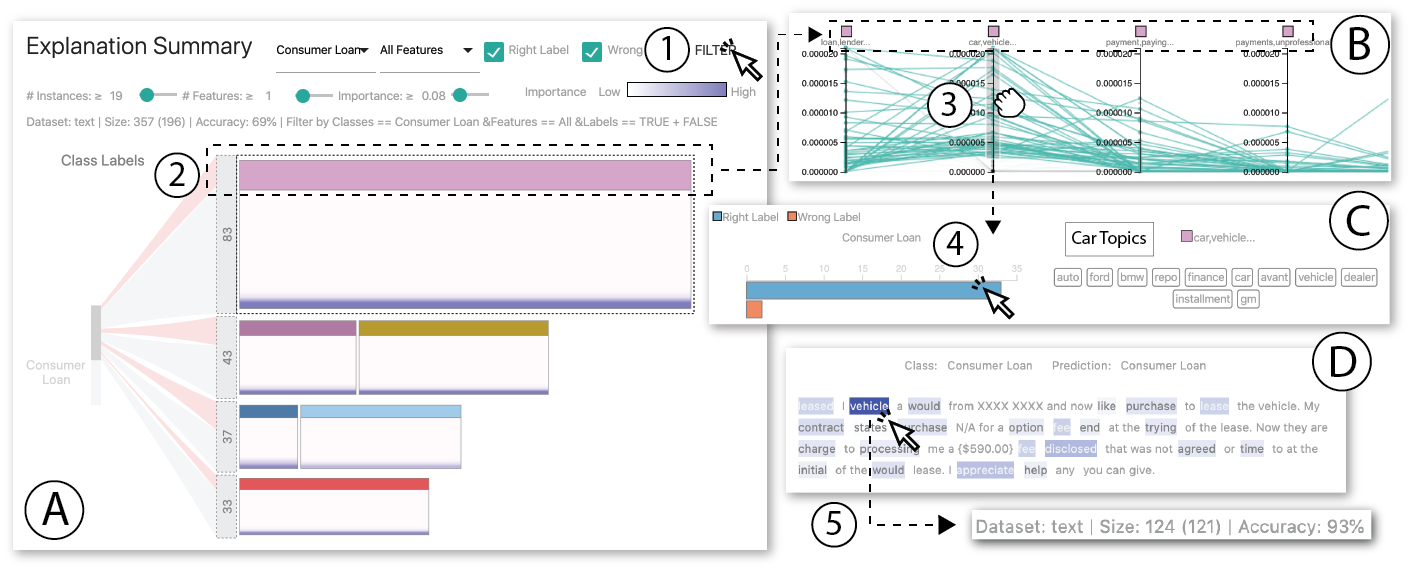}
     \caption{
        Use case of understanding a neural network of document classification trained on text data. A: The explanation summary for customer loan complaints after filtering by class in \clabel{1}. B: By selecting a row cluster in \clabel{2}, the details of explanation values are shown for users to select significant instances and features by brushing in \clabel{3}. C: The subset view displaying the distributions of documents in each class related to the explanatory topics. D: Clicking a blue bar in \clabel{4} shows all the correctly classified documents that are explained by the selected topics. \clabel{5} Clicking the words formulates a query that extracts all documents predicted by the words in the model. 
                }
    \label{fig:case_text}
\end{figure*}

\subsection{Use Cases}
\label{sec:usecase}
We present a usage scenario of understanding deep neural networks related to image recognition and two use cases regarding tabular and text classifications of financial data. Our goal is to demonstrate that our technique generalizes different ML model interpretation challenges in understanding the models and datasets. 

\subsubsection{Usage Scenario: Understanding an Image Classifier}
We first describe a hypothetical walkthrough of understanding what a deep learning model has learned from a set of images (Figure~\ref{fig:teaser}). We use images as examples because the visual presentations are intuitive to understand. Imagine Chris, an ornithologist, wants to study how birds' appearances distinguish their species. He downloads the data and runs the ML model to understand how the machine learns the visual features.

\noindent \textbf{Understand the summary.} Chris uses \algoname to generate an explanation summary consisted of 37 instance clusters and 49 feature clusters. He imports the result to \systemname. After filtering small clusters and clusters with low explanation values, Chris discovers three broad groups of birds with similar prediction logics (Figure~\ref{fig:teaser}\textbf{A}). Each group contains different visual explanatory features (i.e., color blocks), so he decides to go through the instance groups one by one. 

\noindent \textbf{Inspect an interesting subset.} Chris clicks the text box on the row to select all the instances and features from the instance cluster for a detailed inspection. From the subset view (Figure~\ref{fig:teaser}\textbf{B}), he realizes that the neural network learns to group birds with similar colors (yellowish birds) (Figure~\ref{fig:teaser}\clabel{B1}) for a coarse level of decision making. The images then are further classified by more detailed image patches such as the bird's head and belly. Chris notices some classes such as yellow-throated Vimeo have many wrong predictions (images with orange frames) in this subset. Therefore, he clicks on some of the images to examine an image and its classification logics in full detail.

\noindent \textbf{Develop hypotheses by inspecting an instance.} Chris checks an image by clicking it in the subset view. The image and its top explanatory features thus are shown in the instance inspection view (Figure~\ref{fig:teaser}\textbf{C}). He sees a yellow-throated Vimeo is wrongly predicted as a blue-winged warbler because the furs on its neck look similar to the ones of a blue-winged warbler. Chris finds the whole process enjoyable since he quickly identifies the reasoning processes of the model on hundreds of images within a simple journey of visual analysis.

\subsubsection{Tabular Use Case: Understanding the Data Capability}
We now present a use case about approaching the limit of predictability in training a dataset. 
Understanding how the current features help to make predictions allows the financial worker to make improvements to the current credit system.

\noindent \textbf{Understand the summary.}
After filtering by value threshold and the number of instances in the clusters, the analyst obtains a visual explanation summary (Figure~\ref{fig:case_tabular}\textbf{A}). It shows that the blue-colored blocks occupy most of the rows. It consists of mainly items related to delinquency (Figure~\ref{fig:case_tabular}\textbf{B}). Then, he clicks and inspects the subsets and filters some low explanation values by brushing the parallel coordinates (Figure~\ref{fig:case_tabular}\textbf{C}), the subsets show very similar behaviors: for customers who have no history of delinquency, the model labels them as ``good''. 

\noindent \textbf{Discovering more detailed logics in the model.} The analyst sees that such logics provide an approximated accuracy of around $73\%$ in more than half of the population. To understand how a ``bad'' decision is correctly predicted with a good delinquency record, he refines the summary by the classes. The summary shows another logic that influences the model outcome (Figure~\ref{fig:case_tabular}\clabel{A1}). The pink blocks represent the features related to a low external risk estimate, which means that the customer would still be graded as ``bad'' if the external risk estimate is low (Figure~\ref{fig:case_tabular}\clabel{D1}). 

\noindent \textbf{Verifying insights.} From the dataflow, the analyst sees that combining delinquency and risk estimate yields a good prediction result. By verify this hypothesis, he filters the summary by showing only the wrong predictions under the same condition. By adjusting the value threshold to a low extent, the wrong predictions mainly attribute to the fact that they do not have a low risk-estimate (i.e., missing pink blocks related to risk estimate when blue blocks are presented) (Figure~\ref{fig:case_tabular}\clabel{A2}). Clicking the rows with missing pink blocks also reveals that the model fails to identify bad risk when the customer has a good delinquency record and high external risk estimates (Figure~\ref{fig:case_tabular}\clabel{D2}). Throughout the visual analysis in different granularity, the analyst acquires an overview of the model: the model mainly decides by the history of delinquency on the first level of reasoning, then further screen out the bad risks by low external risk estimates. The query panel shows that the rows explained by either these two logics cover more than $70\%$ of the whole dataset.


\subsubsection{Text Use Case: Predicting Customer Complaints}
We present a use case of exploring a text classification model to understand different types of customer complaints. Understanding how customers complain can improve the call center's services. Our financial analyst first uses \algoname to acquire 28 instance clusters and 23 feature clusters. Then, as he is from the loan division in the company, he filters the explanation summary by ``customer loans'' class to explore the customers' inquiries related to loans (Figure~\ref{fig:case_text}(1)).

\noindent\textbf{Identify useful subsets.} The analyst first discovers that the explanation summary is very sparse (Figure~\ref{fig:case_text}\textbf{A}). Therefore, he clicks on the block to examine a more detailed view of the explanation subset (Figure~\ref{fig:case_text}(1)). The details of value distributions in the selected block are shown in the explanation parallel coordinates (Figure~\ref{fig:case_text}\textbf{B}). The analyst discovers that the sparsity mainly comes from the low usage of many topics in the feature clusters. Thus, he brushes the axes of topics that have high values to acquire the subset of instances and topics that heavily correlate to each other. The result of the brushing is shown in the subset view (Figure~\ref{fig:case_text}\textbf{C}).

\noindent\textbf{Discover interesting topics in the subset.} From the subset view, the analyst discovers that many complaints are related to words such as "auto", "bmw", and "ford". These words belong to automobiles and vehicles. Given the longer length of correctly labeled (blue) bars in the bar chart, these words contribute significantly to the correct prediction of customer loan complaints to the model. Therefore, by clicking the blue bar, the analyst inspects the raw documents of these automobile-related instances (Figure~\ref{fig:case_text}\textbf{D}). 

\noindent\textbf{Insights from instances.} By browsing the documents, the analyst confirms that the loan payments complaints are related to vehicle purchases. By clicking the words like ``vehicles'' and ``cars'', he queries the global explanation summary to verify his findings (Figure~\ref{fig:case_text}(5)). The queried results show that there are more than 120 complaints about customer loans that contain such phrases with the correctness of 93\%. Thus, he concludes that automobile purchase is a popular topic when customers approach the financial institution. The company should these topics included during the call center training sessions.

\section{Conclusion and Future Work}
In this work, we present \algoname, an interactive algorithm to construct an explanation summary of an ML model from local explanations of a dataset. The summary allows users to understand the decision rationale and data characteristics together for a holistics XAI experience. With the algorithm, we also present \systemname, an interactive visual analytics system to connect different granularity of XAI tasks. The versatility of our algorithm and system enable scalable visual explorations of generic ML model interpretations on tabular, image, and text data. Our future work includes:

\noindent\textbf{Embed summarization to model training processes.} Instead of generating a summary after the training, we plan to integrate the summary as a layer in the deep neural network to increase the global explanation capability of the model.

\noindent\textbf{User study.} Since many model developers use visualizations such as partial dependency plot or projections to understand the model, we plan to conduct a user study to see if providing explanatory features and similarity among data at the same time will improve any productivity in practice. Also, we plan to conduct a longitudinal evaluation of \systemname to ML researchers to investigate how the system affects model design, data engineering, and model debugging.

\noindent\textbf{Application domain.} Apart from tabular, image, and text classifications, there are also other data primitives such as time series and graph classifications tasks. We plan to explore the visual analytics approach to apply our algorithm to explain ML tasks in these domains.
 

\bibliographystyle{abbrv-doi}
\clearpage
\bibliography{reference}

\begin{thebibliography}{10}

\bibitem{adadi2018peeking}
A.~Adadi and M.~Berrada.
\newblock Peeking inside the black-box: A survey on explainable artificial
  intelligence (xai).
\newblock {\em IEEE Access}, 6:52138--52160, 2018.

\bibitem{amershi2019guidelines}
S.~Amershi, D.~Weld, M.~Vorvoreanu, A.~Fourney, B.~Nushi, P.~Collisson, J.~Suh,
  S.~Iqbal, P.~N. Bennett, K.~Inkpen, et~al.
\newblock Guidelines for human-ai interaction.
\newblock In {\em Proceedings of the 2019 CHI Conference on Human Factors in
  Computing Systems}, pp. 1--13, 2019.

\bibitem{bien2011prototype}
J.~Bien and R.~Tibshirani.
\newblock Prototype selection for interpretable classification.
\newblock {\em The Annals of Applied Statistics}, pp. 2403--2424, 2011.

\bibitem{bilal2017convolutional}
A.~Bilal, A.~Jourabloo, M.~Ye, X.~Liu, and L.~Ren.
\newblock Do convolutional neural networks learn class hierarchy?
\newblock {\em IEEE transactions on visualization and computer graphics},
  24(1):152--162, 2017.

\bibitem{boukhelifa2017data}
N.~Boukhelifa, M.-E. Perrin, S.~Huron, and J.~Eagan.
\newblock How data workers cope with uncertainty: A task characterisation
  study.
\newblock In {\em Proceedings of the 2017 CHI Conference on Human Factors in
  Computing Systems}, pp. 3645--3656, 2017.

\bibitem{carvalho2019machine}
D.~V. Carvalho, E.~M. Pereira, and J.~S. Cardoso.
\newblock Machine learning interpretability: A survey on methods and metrics.
\newblock {\em Electronics}, 8(8):832, 2019.

\bibitem{chandola2007summarization}
V.~Chandola and V.~Kumar.
\newblock Summarization--compressing data into an informative representation.
\newblock {\em Knowledge and Information Systems}, 12(3):355--378, 2007.

\bibitem{charikar2002similarity}
M.~S. Charikar.
\newblock Similarity estimation techniques from rounding algorithms.
\newblock In {\em Proceedings of the thiry-fourth annual ACM symposium on
  Theory of computing}, pp. 380--388, 2002.

\bibitem{chen2019looks}
C.~Chen, O.~Li, D.~Tao, A.~Barnett, C.~Rudin, and J.~K. Su.
\newblock This looks like that: deep learning for interpretable image
  recognition.
\newblock In {\em Advances in Neural Information Processing Systems}, pp.
  8928--8939, 2019.

\bibitem{craven1996extracting}
M.~Craven and J.~W. Shavlik.
\newblock Extracting tree-structured representations of trained networks.
\newblock In {\em Advances in neural information processing systems}, pp.
  24--30, 1996.

\bibitem{dhillon2001co}
I.~S. Dhillon.
\newblock Co-clustering documents and words using bipartite spectral graph
  partitioning.
\newblock In {\em Proceedings of the seventh ACM SIGKDD international
  conference on Knowledge discovery and data mining}, pp. 269--274, 2001.

\bibitem{dhillon2003information}
I.~S. Dhillon, S.~Mallela, and D.~S. Modha.
\newblock Information-theoretic co-clustering.
\newblock In {\em Proceedings of the ninth ACM SIGKDD international conference
  on Knowledge discovery and data mining}, pp. 89--98, 2003.

\bibitem{ghoniem2005readability}
M.~Ghoniem, J.-D. Fekete, and P.~Castagliola.
\newblock On the readability of graphs using node-link and matrix-based
  representations: a controlled experiment and statistical analysis.
\newblock {\em Information Visualization}, 4(2):114--135, 2005.

\bibitem{gilpin2018explaining}
L.~H. Gilpin, D.~Bau, B.~Z. Yuan, A.~Bajwa, M.~Specter, and L.~Kagal.
\newblock Explaining explanations: An overview of interpretability of machine
  learning.
\newblock In {\em 2018 IEEE 5th International Conference on data science and
  advanced analytics (DSAA)}, pp. 80--89. IEEE, 2018.

\bibitem{grover2016node2vec}
A.~Grover and J.~Leskovec.
\newblock node2vec: Scalable feature learning for networks.
\newblock In {\em Proceedings of the 22nd ACM SIGKDD international conference
  on Knowledge discovery and data mining}, pp. 855--864, 2016.

\bibitem{guidotti2018survey}
R.~Guidotti, A.~Monreale, S.~Ruggieri, F.~Turini, F.~Giannotti, and
  D.~Pedreschi.
\newblock A survey of methods for explaining black box models.
\newblock {\em ACM computing surveys (CSUR)}, 51(5):1--42, 2018.

\bibitem{hlawatsch2014visual}
M.~Hlawatsch, M.~Burch, and D.~Weiskopf.
\newblock Visual adjacency lists for dynamic graphs.
\newblock {\em IEEE transactions on visualization and computer graphics},
  20(11):1590--1603, 2014.

\bibitem{hohman2019gamut}
F.~Hohman, A.~Head, R.~Caruana, R.~DeLine, and S.~M. Drucker.
\newblock Gamut: A design probe to understand how data scientists understand
  machine learning models.
\newblock In {\em Proceedings of the 2019 CHI Conference on Human Factors in
  Computing Systems}, pp. 1--13, 2019.

\bibitem{hohman2018visual}
F.~Hohman, M.~Kahng, R.~Pienta, and D.~H. Chau.
\newblock Visual analytics in deep learning: An interrogative survey for the
  next frontiers.
\newblock {\em IEEE transactions on visualization and computer graphics},
  25(8):2674--2693, 2018.

\bibitem{hohman2019s}
F.~Hohman, H.~Park, C.~Robinson, and D.~H.~P. Chau.
\newblock Summit: Scaling deep learning interpretability by visualizing
  activation and attribution summarizations.
\newblock {\em IEEE transactions on visualization and computer graphics},
  26(1):1096--1106, 2019.

\bibitem{holstein2019improving}
K.~Holstein, J.~Wortman~Vaughan, H.~Daum{\'e}~III, M.~Dudik, and H.~Wallach.
\newblock Improving fairness in machine learning systems: What do industry
  practitioners need?
\newblock In {\em Proceedings of the 2019 CHI Conference on Human Factors in
  Computing Systems}, pp. 1--16, 2019.

\bibitem{kahng2017cti}
M.~Kahng, P.~Y. Andrews, A.~Kalro, and D.~H.~P. Chau.
\newblock Activis: Visual exploration of industry-scale deep neural network
  models.
\newblock {\em IEEE transactions on visualization and computer graphics},
  24(1):88--97, 2017.

\bibitem{kim2014bayesian}
B.~Kim, C.~Rudin, and J.~A. Shah.
\newblock The bayesian case model: A generative approach for case-based
  reasoning and prototype classification.
\newblock In {\em Advances in Neural Information Processing Systems}, pp.
  1952--1960, 2014.

\bibitem{kim2018interpretability}
B.~Kim, M.~Wattenberg, J.~Gilmer, C.~Cai, J.~Wexler, F.~Viegas, et~al.
\newblock Interpretability beyond feature attribution: Quantitative testing
  with concept activation vectors (tcav).
\newblock In {\em International Conference on Machine Learning}, pp.
  2668--2677, 2018.

\bibitem{li2018deep}
O.~Li, H.~Liu, C.~Chen, and C.~Rudin.
\newblock Deep learning for case-based reasoning through prototypes: A neural
  network that explains its predictions.
\newblock In {\em Thirty-Second AAAI Conference on Artificial Intelligence},
  2018.

\bibitem{li2018embeddingvis}
Q.~Li, K.~S. Njotoprawiro, H.~Haleem, Q.~Chen, C.~Yi, and X.~Ma.
\newblock Embeddingvis: A visual analytics approach to comparative network
  embedding inspection.
\newblock In {\em 2018 IEEE Conference on Visual Analytics Science and
  Technology (VAST)}, pp. 48--59. IEEE, 2018.

\bibitem{liao2020questioning}
Q.~V. Liao, D.~Gruen, and S.~Miller.
\newblock Questioning the ai: Informing design practices for explainable ai
  user experiences.
\newblock {\em arXiv preprint arXiv:2001.02478}, 2020.

\bibitem{liu2018analyzing}
M.~Liu, S.~Liu, H.~Su, K.~Cao, and J.~Zhu.
\newblock Analyzing the noise robustness of deep neural networks.
\newblock In {\em 2018 IEEE Conference on Visual Analytics Science and
  Technology (VAST)}, pp. 60--71. IEEE, 2018.

\bibitem{liu2016towards}
M.~Liu, J.~Shi, Z.~Li, C.~Li, J.~Zhu, and S.~Liu.
\newblock Towards better analysis of deep convolutional neural networks.
\newblock {\em IEEE transactions on visualization and computer graphics},
  23(1):91--100, 2016.

\bibitem{liu2017visual}
S.~Liu, P.-T. Bremer, J.~J. Thiagarajan, V.~Srikumar, B.~Wang, Y.~Livnat, and
  V.~Pascucci.
\newblock Visual exploration of semantic relationships in neural word
  embeddings.
\newblock {\em IEEE transactions on visualization and computer graphics},
  24(1):553--562, 2017.

\bibitem{lundberg2017unified}
S.~M. Lundberg and S.-I. Lee.
\newblock A unified approach to interpreting model predictions.
\newblock In {\em Advances in neural information processing systems}, pp.
  4765--4774, 2017.

\bibitem{martens2008decompositional}
D.~Martens, B.~Baesens, and T.~Van~Gestel.
\newblock Decompositional rule extraction from support vector machines by
  active learning.
\newblock {\em IEEE Transactions on Knowledge and Data Engineering},
  21(2):178--191, 2008.

\bibitem{ming2017understanding}
Y.~Ming, S.~Cao, R.~Zhang, Z.~Li, Y.~Chen, Y.~Song, and H.~Qu.
\newblock Understanding hidden memories of recurrent neural networks.
\newblock In {\em 2017 IEEE Conference on Visual Analytics Science and
  Technology (VAST)}, pp. 13--24. IEEE, 2017.

\bibitem{ming2018rulematrix}
Y.~Ming, H.~Qu, and E.~Bertini.
\newblock Rulematrix: Visualizing and understanding classifiers with rules.
\newblock {\em IEEE transactions on visualization and computer graphics},
  25(1):342--352, 2018.

\bibitem{ming2019protosteer}
Y.~Ming, P.~Xu, F.~Cheng, H.~Qu, and L.~Ren.
\newblock Protosteer: Steering deep sequence model with prototypes.
\newblock {\em IEEE Transactions on Visualization and Computer Graphics},
  26(1):238--248, 2019.

\bibitem{ming2019interpretable}
Y.~Ming, P.~Xu, H.~Qu, and L.~Ren.
\newblock Interpretable and steerable sequence learning via prototypes.
\newblock In {\em Proceedings of the 25th ACM SIGKDD International Conference
  on Knowledge Discovery \& Data Mining}, pp. 903--913, 2019.

\bibitem{mohseni2018survey}
S.~Mohseni, N.~Zarei, and E.~D. Ragan.
\newblock A survey of evaluation methods and measures for interpretable machine
  learning.
\newblock {\em arXiv preprint arXiv:1811.11839}, 2018.

\bibitem{muhlbacher2017treepod}
T.~M{\"u}hlbacher, L.~Linhardt, T.~M{\"o}ller, and H.~Piringer.
\newblock Treepod: Sensitivity-aware selection of pareto-optimal decision
  trees.
\newblock {\em IEEE transactions on visualization and computer graphics},
  24(1):174--183, 2017.

\bibitem{muller2019data}
M.~Muller, I.~Lange, D.~Wang, D.~Piorkowski, J.~Tsay, Q.~V. Liao, C.~Dugan, and
  T.~Erickson.
\newblock How data science workers work with data: Discovery, capture,
  curation, design, creation.
\newblock In {\em Proceedings of the 2019 CHI Conference on Human Factors in
  Computing Systems}, pp. 1--15, 2019.

\bibitem{navlakha2008graph}
S.~Navlakha, R.~Rastogi, and N.~Shrivastava.
\newblock Graph summarization with bounded error.
\newblock In {\em Proceedings of the 2008 ACM SIGMOD international conference
  on Management of data}, pp. 419--432, 2008.

\bibitem{okoe2018node}
M.~Okoe, R.~Jianu, and S.~Kobourov.
\newblock Node-link or adjacency matrices: Old question, new insights.
\newblock {\em IEEE transactions on visualization and computer graphics},
  25(10):2940--2952, 2018.

\bibitem{pezzotti2017deepeyes}
N.~Pezzotti, T.~H{\"o}llt, J.~Van~Gemert, B.~P. Lelieveldt, E.~Eisemann, and
  A.~Vilanova.
\newblock Deepeyes: Progressive visual analytics for designing deep neural
  networks.
\newblock {\em IEEE transactions on visualization and computer graphics},
  24(1):98--108, 2017.

\bibitem{poerner2018evaluating}
N.~Poerner, B.~Roth, and H.~Sch{\"u}tze.
\newblock Evaluating neural network explanation methods using hybrid documents
  and morphological agreement.
\newblock {\em arXiv preprint arXiv:1801.06422}, 2018.

\bibitem{ras2018explanation}
G.~Ras, M.~van Gerven, and P.~Haselager.
\newblock Explanation methods in deep learning: Users, values, concerns and
  challenges.
\newblock In {\em Explainable and Interpretable Models in Computer Vision and
  Machine Learning}, pp. 19--36. Springer, 2018.

\bibitem{rauber2016visualizing}
P.~E. Rauber, S.~G. Fadel, A.~X. Falcao, and A.~C. Telea.
\newblock Visualizing the hidden activity of artificial neural networks.
\newblock {\em IEEE transactions on visualization and computer graphics},
  23(1):101--110, 2016.

\bibitem{ribeiro2016should}
M.~T. Ribeiro, S.~Singh, and C.~Guestrin.
\newblock " why should i trust you?" explaining the predictions of any
  classifier.
\newblock In {\em Proceedings of the 22nd ACM SIGKDD international conference
  on knowledge discovery and data mining}, pp. 1135--1144, 2016.

\bibitem{rule2018exploration}
A.~Rule, A.~Tabard, and J.~D. Hollan.
\newblock Exploration and explanation in computational notebooks.
\newblock In {\em Proceedings of the 2018 CHI Conference on Human Factors in
  Computing Systems}, pp. 1--12, 2018.

\bibitem{satopaa2011finding}
V.~Satopaa, J.~Albrecht, D.~Irwin, and B.~Raghavan.
\newblock Finding a" kneedle" in a haystack: Detecting knee points in system
  behavior.
\newblock In {\em 2011 31st international conference on distributed computing
  systems workshops}, pp. 166--171. IEEE, 2011.

\bibitem{sawada2019model}
S.~Sawada and M.~Toyoda.
\newblock Model-agnostic visual explanation of machine learning models based on
  heat map.
\newblock 2019.

\bibitem{shrikumar2017learning}
A.~Shrikumar, P.~Greenside, and A.~Kundaje.
\newblock Learning important features through propagating activation
  differences.
\newblock In {\em Proceedings of the 34th International Conference on Machine
  Learning-Volume 70}, pp. 3145--3153. JMLR. org, 2017.

\bibitem{simonyan2014deep}
K.~Simonyan, A.~Vedaldi, and A.~Zisserman.
\newblock Deep inside convolutional networks: Visualising image classification
  models and saliency maps.
\newblock In {\em Workshop at International Conference on Learning
  Representations}, 2014.

\bibitem{strobelt2018s}
H.~Strobelt, S.~Gehrmann, M.~Behrisch, A.~Perer, H.~Pfister, and A.~M. Rush.
\newblock S eq 2s eq-v is: A visual debugging tool for sequence-to-sequence
  models.
\newblock {\em IEEE transactions on visualization and computer graphics},
  25(1):353--363, 2018.

\bibitem{strobelt2017lstmvis}
H.~Strobelt, S.~Gehrmann, H.~Pfister, and A.~M. Rush.
\newblock Lstmvis: A tool for visual analysis of hidden state dynamics in
  recurrent neural networks.
\newblock {\em IEEE transactions on visualization and computer graphics},
  24(1):667--676, 2017.

\bibitem{sundararajan2017axiomatic}
M.~Sundararajan, A.~Taly, and Q.~Yan.
\newblock Axiomatic attribution for deep networks.
\newblock In {\em Proceedings of the 34th International Conference on Machine
  Learning-Volume 70}, pp. 3319--3328. JMLR. org, 2017.

\bibitem{tzeng2005opening}
F.~. {Tzeng} and K.~. {Ma}.
\newblock Opening the black box - data driven visualization of neural networks.
\newblock In {\em VIS 05. IEEE Visualization, 2005.}, pp. 383--390, Oct 2005.
  doi: {{%
10\hspace{.1pt}\discretionary{.}{%
}{.}\hspace{.4pt}1109\discretionary{/}{%
}{/}VISUAL\hspace{.1pt}\discretionary{.}{%
}{.}\hspace{.4pt}2005\hspace{.1pt}\discretionary{.}{%
}{.}\hspace{.4pt}1532820}}


\bibitem{van2011baobabview}
S.~Van Den~Elzen and J.~J. van Wijk.
\newblock Baobabview: Interactive construction and analysis of decision trees.
\newblock In {\em 2011 IEEE conference on visual analytics science and
  technology (VAST)}, pp. 151--160. IEEE, 2011.

\bibitem{voigt2017eu}
P.~Voigt and A.~Von~dem Bussche.
\newblock The eu general data protection regulation (gdpr).
\newblock {\em A Practical Guide, 1st Ed., Cham: Springer International
  Publishing}, 2017.

\bibitem{wexler2019if}
J.~Wexler, M.~Pushkarna, T.~Bolukbasi, M.~Wattenberg, F.~Vi{\'e}gas, and
  J.~Wilson.
\newblock The what-if tool: Interactive probing of machine learning models.
\newblock {\em IEEE transactions on visualization and computer graphics},
  26(1):56--65, 2019.

\bibitem{wongsuphasawat2017visualizing}
K.~Wongsuphasawat, D.~Smilkov, J.~Wexler, J.~Wilson, D.~Mane, D.~Fritz,
  D.~Krishnan, F.~B. Vi{\'e}gas, and M.~Wattenberg.
\newblock Visualizing dataflow graphs of deep learning models in tensorflow.
\newblock {\em IEEE transactions on visualization and computer graphics},
  24(1):1--12, 2017.

\bibitem{wu2019errudite}
T.~Wu, M.~T. Ribeiro, J.~Heer, and D.~S. Weld.
\newblock Errudite: Scalable, reproducible, and testable error analysis.
\newblock In {\em Proceedings of the 57th Annual Meeting of the Association for
  Computational Linguistics}, pp. 747--763, 2019.

\bibitem{xiang2020interactive}
S.~{Xiang}, X.~{Ye}, J.~{Xia}, J.~{Wu}, Y.~{Chen}, and S.~{Liu}.
\newblock Interactive correction of mislabeled training data.
\newblock In {\em 2019 IEEE Conference on Visual Analytics Science and
  Technology (VAST)}, pp. 57--68, Oct 2019. doi: {{%
10\hspace{.1pt}\discretionary{.}{%
}{.}\hspace{.4pt}1109\discretionary{/}{%
}{/}VAST47406\hspace{.1pt}\discretionary{.}{%
}{.}\hspace{.4pt}2019\hspace{.1pt}\discretionary{.}{%
}{.}\hspace{.4pt}8986943}}


\bibitem{yang2017scalable}
H.~Yang, C.~Rudin, and M.~Seltzer.
\newblock Scalable bayesian rule lists.
\newblock In {\em Proceedings of the 34th International Conference on Machine
  Learning-Volume 70}, pp. 3921--3930. JMLR. org, 2017.

\bibitem{zeiler2014visualizing}
M.~D. Zeiler and R.~Fergus.
\newblock Visualizing and understanding convolutional networks.
\newblock In {\em European conference on computer vision}, pp. 818--833.
  Springer, 2014.

\bibitem{zhao2018iforest}
X.~Zhao, Y.~Wu, D.~L. Lee, and W.~Cui.
\newblock iforest: Interpreting random forests via visual analytics.
\newblock {\em IEEE transactions on visualization and computer graphics},
  25(1):407--416, 2018.

\end{thebibliography}
\clearpage
\begin{figure*}
    \centering
     \includegraphics[width=\linewidth]{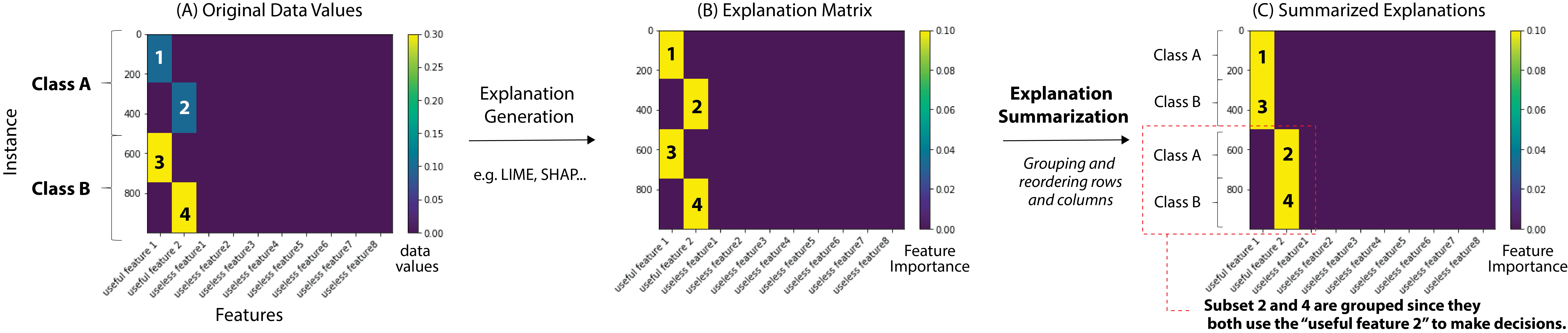}
     \caption{
        A workflow of ML model explanation generation process as well as our summarization workflow. A: The original data with its values that are used by the ML models to make decisions. B: After the ML model is trained, the explanation model generates the explanation values, in other words feature importance, for each instance and its features.
        C: Our goal is to group the instances and features that co-exist in the decision making process. For example, the portion of subset ``2'' and ``4'' are grouped in the final summary output since it conveys an insight that these portion of data distinguish the differences between the classes using mainly the ``useful feature 2'' in the original data.
                }
    \label{fig:synthetic_workflow}
\end{figure*}
\section{Appendix}
\label{sec:appendix}

\subsection{Background of Local Explanation Models}
\label{sec:background}
We provide a background of the mainstream models that generate local explanations of a machine learning model's decisions to a dataset.
The popularity of giving local explanations, except applying logical models such as decision trees or rules, is because these methods provide an independent and highly customized explanation for each instance. When explanations do not aggregate into general decisions or rules, they become more faithful to the original model.

In general, to generate a local explanation for an instance, explanation algorithms usually seek one of the following approaches:

\begin{enumerate}
    \item \textit{Local Linear Models}: The algorithm searches the neighbors of an instance, then fits the subset to a linear model such that the higher the gradient of a feature in the linear model, the more important the feature is to the prediction of the selected instances. SHAP \cite{lundberg2017unified} and LIME \cite{ribeiro2016should} are the examples that use neighbors to evaluate an instance.

    \item \textit{Perturbation}: Instead of using other instances to generate explanations, one can perturbate the values of its attributes and observe whether the output changes significantly after removing, masking or altering them. The sensitivity of each feature implies that its value lies in the decision boundary of the machine learning model. Thus, a sensitive feature from perturbation has a high influential power on the instance. This method has been applied to Convolutional Neural Networks (CNNs) for image classification \cite{zeiler2014visualizing}.
    
    \item \textit{Prototype}: The intuition is to use representative original training data (i.e. prototype) to explain a prediction, which can be selected by clustering the latent representations in the ML model. Given the class labels of each prototype and their similarities with the input data, the prediction and reasoning process becomes a scoring system where the class with the highest score (i.e. the sum of similarities of prototypes belonging to the class) is the returned result. This technique has been readily incorporated in deep neural networks for image, text and sequence predictions \cite{chen2019looks,li2018deep,ming2019interpretable}.

    \item \textit{Backpropagation}: Since complex models like neural networks contain series of propagation of weights from the input to the output neurons to produce predictions, one can invert the process to backpropagate the neurons with great gradients from the output to the input data locate the portion of original data that causes the neuron activations in the output. Such a portion implies the meaningful features that explain the model's decision. Saliency Maps \cite{simonyan2014deep}, DeepLIFT \cite{shrikumar2017learning} and Intgrad \cite{sundararajan2017axiomatic} are the examples of such methods.
\end{enumerate}

\subsection{Tasks Breakdown of Explainable AI}
\label{sec:tasks_appendix}
In general, ML models explainability is achieved by three different types of tasks \textbf{(T)}: 
\textbf{Global} (\textit{general behavior of a model}), \textbf{Local} (\textit{behavior of a model to an instance}), and \textbf{Class} (\textit{behavior of a model to a class}) \cite{guidotti2018survey,liao2020questioning}.
Liao \textit{et.al.} \cite{liao2020questioning} in addition provides actionable suggestions \textbf{(A)} for each task.
For each task and action, we identify different opportunities (\textbf{O}) that an explanation summary can help achieve the tasks.
\begin{itemize}[noitemsep]
	\item[\textbf{T.1}]\textbf{Global Explanation}. 
	The goal is to understand the overall weights of features used by the model to explain how AI makes decisions on the dataset in general.
	\begin{itemize}[noitemsep,topsep=0pt,leftmargin=6mm]
		\item[\textbf{A.1}] 
		Users \textit{select the important features} that affect the whole dataset's outcome
		to uncover data-related issues such as data collection, bias, and privacy. 
		\item[\textbf{A.2}] They may also \textit{evaluate the limit and capability of the model}. 
		By inspecting the main features, users can develop a mental model to interact with or improve the system.
	\end{itemize} 
	\begin{itemize}[noitemsep,topsep=0pt,leftmargin=6mm]
		\item[\textbf{O.1}] 
		An explanation summary can define the appropriate \textit{level of details} 
		to explain the model without losing too many details nor overwhelming the users.
		Grouping the relevant features and instances allow interactions for users to prioritize different information shown at a time.
		\item[\textbf{O.2}] 
		Subsetting the information by similarity also decreases the complexity of the explanation since 
		the instances and features shown have many common properties. 
		This allows the global explanations visualized to be more \textit{representative}.
	\end{itemize}
	\item[\textbf{T.2}] \textbf{Local Explanation}. 
	The goal here is to inspect the model's behavior on a specific instance and understand how the instance's properties influence the outcome.
	\begin{itemize}[noitemsep,topsep=0pt,leftmargin=6mm]
		\item[\textbf{A.3}] 
		A popular activity is to explore different \textit{what-if} scenarios.
		Users observe the outcome if some features become different
		which helps to explore more scenarios of applying the model and gain insights into the model's capability.
		\item[\textbf{A.4}] 
		Another action is to directly understand \textit{why} does the instance belong to a prediction and \textit{why not} does it result in other outcomes.
		This helps to discover the local decision boundaries of the model.
		\item[\textbf{A.5}] 
		Providing the \textit{original input/data} provides a more holistic system capability to understand a particular decision and accommodate users' understandings and interactions. 
	\end{itemize} 
	\begin{itemize}[noitemsep,topsep=0pt,leftmargin=6mm]
		\item[\textbf{O.3}] 
		Grouping similar instances provide \textit{neighbors of the instance} that are explained similarly by the model,
		which increases the number of instances to support users' insights and findings.
		Can we conclude that ``ears'' are important in the prediction of "cats" from what we see on a single image?
		We also know that if there exists lots of cat images exhibiting similar characteristics. 
		An explanation summary thus allows a large set of instances to be analyzed to avoid spurious conclusions \cite{wu2019errudite}. 
		\item[\textbf{O.4}] 
		Similar to grouping instances, grouping features allows users to \textit{prioritize important features} that explain an instance and its neighbors,
		which reduce the cognitive workload when deriving understandings to the model's decision logic.
		
	\end{itemize}
	\item[\textbf{T.3}] \textbf{Class Explanation} (\textit{Counterfactual}). 
	How a prediction (class) works in the model is also an important emphasis.
	It is similar to a global explanation but with a smaller granularity on a specific class.
	Yet, the actions to understand a class are more similar to instance explanations, which focus on the sensitivity of features to each prediction. 
	\begin{itemize}[noitemsep,topsep=0pt,leftmargin=6mm]
		\item[\textbf{A.6}]
		Testing the sensitivity of features towards a prediction is equivalent to the test of different \textit{what-if} scenarios.
		By testing different ranges of features, users can understand the decision boundaries of a prediction class.
		\item[\textbf{A.7}] 
		Besides interaction, the exploration of the relevant features of a prediction also helps 
		understand \textit{why} and \textit{why not} cases of a prediction to gain insights into the decision logic.
		\item[\textbf{O.5}] 
        With groups of similar instances and features, users can apply different \textit{levels of details} to acquire more precise subsets.
        \item[\textbf{O.6}] 
		Extending the findings of an instance to its similar neighbors inside a class increases the confidences of the insights.
	\end{itemize} 	 
\end{itemize}

\subsection{End-to-End Explanation Modeling Pipeline}
\label{sec:pipeline}
In this section, we describe the example pipelines in handling tabular, image, and text data that result in explanation matrices with the explanatory features in Table~\ref{table:nature}. We explicitly categorize the pipeline with \textbf{preprocessing}, \textbf{ML modeling}, and \textbf{explanation modeling} stages. Notice that they are not the only ways to achieve the objectives of data engineering. The explanation models can be interchanged as well. In addition, we provide an synthetic example to illustrate how the whole explanation process works, as well as our goal to summarize the whole explanation data in Figure~\ref{fig:synthetic_workflow}.

\subsubsection{Tabular Data}
\textbf{Preprocessing.} To enable logics as the explanatory features for tabular data, we need to preprocess the original data into one-hot encodings of logics under each attribute. For numerical and ordinal data, the attributes are first discretized into different quantiles. Then, the one-hot encoding can be applied to transform the quantiles into separate columns, where 0 indicates the data does not fall into the ranges while 1 indicates it does. The way of discretizing attributes can be as straightforward as choosing a fixed number of equal intervals or leveraging the statistical properties such as entropy. In our use case, we use Sturge's rule to determine the number of quantiles and the ranges of quantiles are determined by the training data. The one-hot encoding can also directly be applied to categorical attributes. 

\noindent\textbf{ML modeling.} Then, the transformed data is used to train a neural network so that the logics are the input features. This allows the logic to be evaluated in the explanation methods. 

\noindent\textbf{Explanation Modeling.} As the input features of the ML model are a set of logics, methods such as LIME and SHAP can be directly applied to the model and dataset to generate feature vectors composed of a set of logics.

\subsubsection{Images}
\textbf{Preprocessing.} For images, we do not need much feature engineering as the explanatory features are the pixels themselves. We only need to apply standard image augmenting techniques (i.e. replicating training images with scaling, rotating, and mirroring) to increase the training data size for a better model accuracy.

\noindent\textbf{ML modeling.} We apply prototype learning inside a Convolutional Neural Network \cite{chen2019looks}. It adds a prototype layer on the last layer of the original neural network. The training process results in a selection of a fixed number of image patches from the training data as prototypes that are used to reason the prediction of new data. 

\noindent\textbf{Explanation Modeling.} As the explanation model is already incorporated as a layer in the ML model when new data comes in, an $n \times m$ explanation matrix can be constructed, where $n$ is the number of tested data and $m$ is the number of prototypes.

\subsubsection{Text}
\textbf{Preprocessing.} Similar to images, the explanation of text comes from the texts inside the documents as well. Thus, we only need to apply standard text preprocessing steps like removing stopwords and infrequent words to make sure the explanation models do not return explanations with meaningless topics. 

\noindent\textbf{ML modeling.} We can use common text models such as RNN and LSTM to generate predictions. Notice that usually the first layer of these models are the word embeddings of the whole dataset. We can leverage this word embeddings to find extract the topics in the dataset by clustering based on them.

\noindent\textbf{Explanation Modeling.} For training the ML model, we can examine each word's importance to the prediction by gradient-based explanation models such as DeepLIFT and Intgrad. This results in an extremely sparse matrix where each feature is a word that appears in more or equal than one documents. Also, words with similar meanings such as ``good'' and ``excellent'' will be treated as different features. To densify the explanation matrix so that similar words are grouped and more significant hidden structures can be produced, we can transform the local explanation from a feature vector of words to a feature vector of topics. The explanation importance of each topic to an instance can be determined by the maximum explanation importance among the words in the topic. Such allows words with similar semantics to be grouped before the matrix is summarized.

\begin{figure*}
    \centering
     \includegraphics[width=\linewidth]{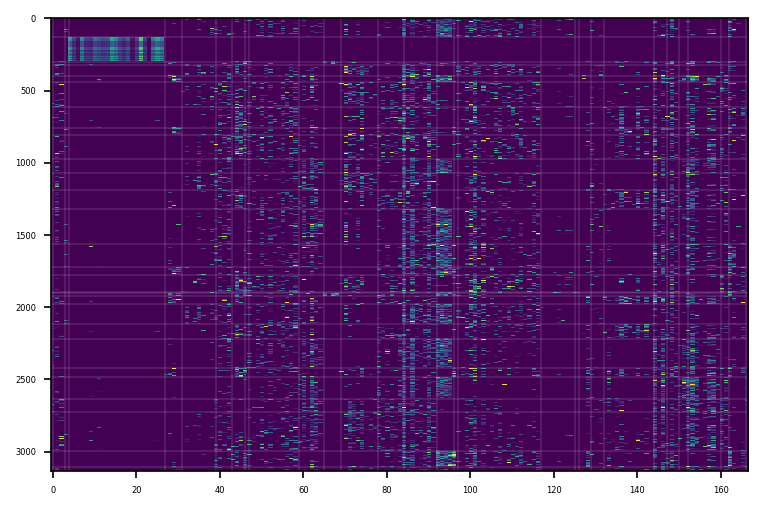}
     \caption{
       Explanation summary matrix of the HELOC dataset's explanation matrix. Rows represent the instances and columns represent the explanatory features. Vertical lines represent row clusters and horizontal lines represent column clusters. The color reflects the explanation values.
                }
    \label{fig:tabular_matrix}
\end{figure*}

\begin{figure*}
    \centering
     \includegraphics[width=\linewidth]{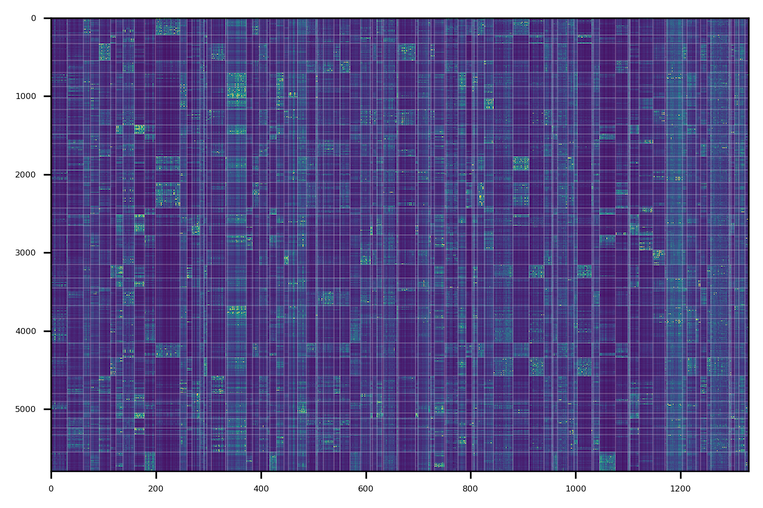}
     \caption{
        Explanation summary matrix of the Caltech-UCSD Birds-200-2011 Images' explanation matrix. Rows represent the instances and columns represent the explanatory features. Vertical lines represent row clusters and horizontal lines represent column clusters. The color reflects the explanation values.
                }
    \label{fig:image_matrix}
\end{figure*}

\begin{figure*}
    \centering
     \includegraphics[width=\linewidth]{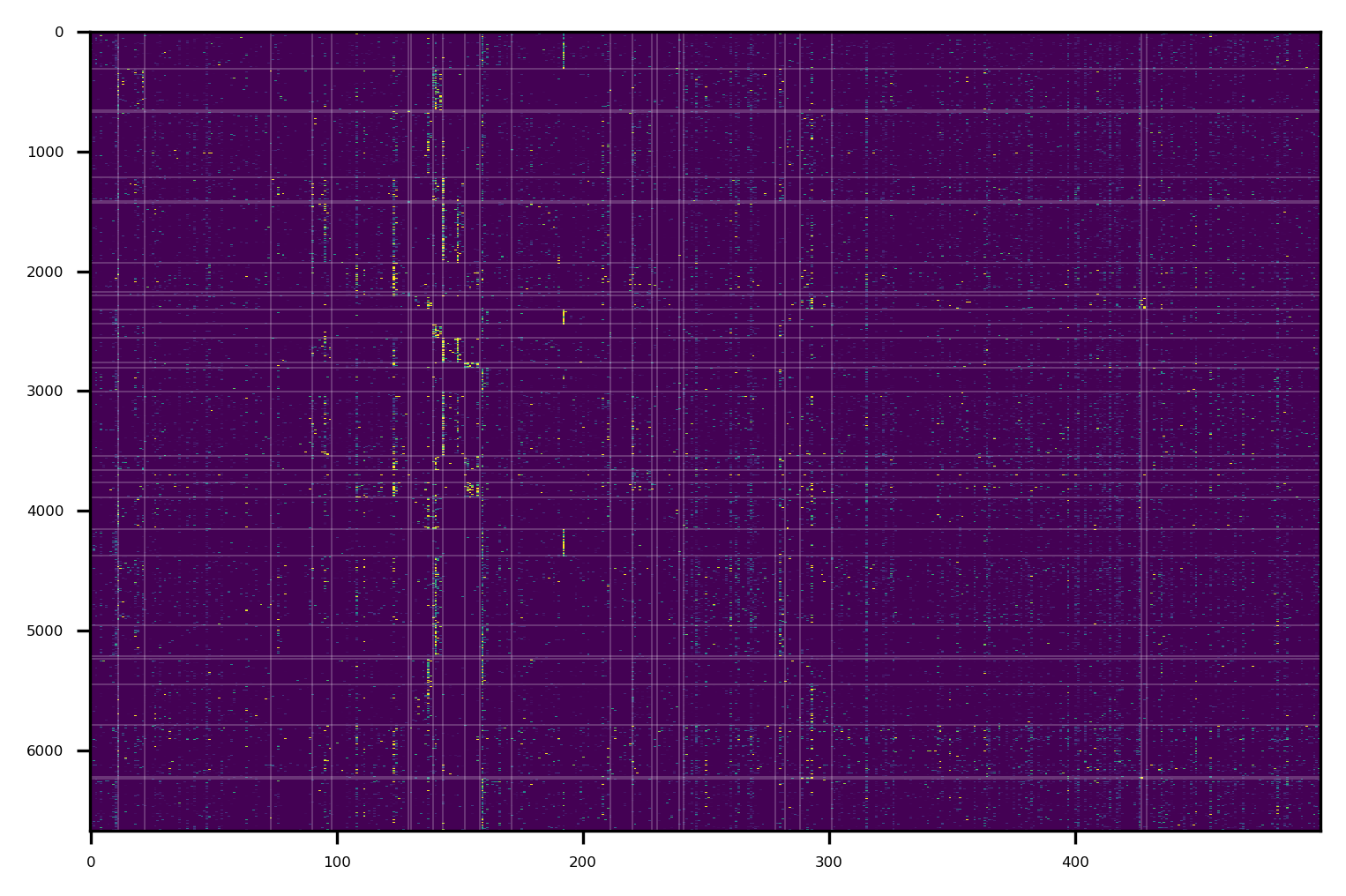}
     \caption{
        Explanation summary matrix of the US Consumer Finance Complaints' explanation matrix. Rows represent the instances and columns represent the explanatory features. Vertical lines represent row clusters and horizontal lines represent column clusters. The color reflects the explanation values.
                }
    \label{fig:text_matrix}
\end{figure*}
\end{document}